\documentclass[useAMS,usenatbib]{mn2e}
\usepackage{natbib}
\usepackage{graphicx}
\usepackage{times}
\usepackage{subfigure}
\usepackage{url}
\usepackage{amsmath}
\usepackage{amssymb}
\usepackage{longtable,lscape}
\usepackage{cite}
\usepackage{color}
\usepackage{placeins}
\usepackage{float}
\usepackage{gensymb}
\usepackage{hyperref}
\def\hii{\mbox{H{\sc ii}}}
\def\feii{\mbox{[Fe{\sc ii}]}}
\def\h2{H$_2$}
\def\tab{Table~}
\def\fig{Fig.~}
\def\g19{G19.88-0.53}

\def\um{$\rm \mu m$}
\def\arc{$\arcsec$}

\title[Extended green object G19.88-0.53]{Multiwavelength investigation of extended green object G19.88-0.53: Revealing a protocluster}

\author[Issac et al.]{Namitha Issac$^1$\thanks{E-mail: namithaissac.16@res.iist.ac.in}, Anandmayee Tej$^1$, Tie Liu$^{2,3}$, Watson Varricatt$^4$, Sarita Vig$^1$
\newauthor 
Ishwara Chandra C.H.$^5$, Mathias Schultheis$^6$, Govind Nandakumar$^{7,8}$ \\
$^1$Indian Institute of Space Science and Technology, Thiruvananthapuram 695 547, Kerala, India\\
$^2$Shanghai Astronomical Observatory, Chinese Academy of Sciences, 80 Nandan Road, Shanghai 200030, People’s Republic of China \\
$^3$Key Laboratory for Research in Galaxies and Cosmology, Chinese Academy of Sciences, 80 Nandan Road, Shanghai 200030, People’s Republic of China \\
$^4$Institute for Astronomy, UKIRT Observatory, 660 N. A\' ohoku place, Hilo, HI 96720, USA \\
$^5$National Centre for Radio Astrophysics (NCRA-TIFR), Pune, India \\
$^6$Laboratoire Lagrange, Universit\' e C\^ote d'Azur, Observatoire de la C\^ote d'Azur, CNRS, Blvd de l'Observatoire, F-06304 Nice, France \\
$^7$Research School of Astronomy \& Astrophysics, Australian National University, ACT 2611, Australia \\
$^8$ARC Centre of Excellence for All Sky Astrophysics in Three Dimensions (ASTRO-3D), Australia \\}

\begin{document}

\date{}

\pagerange{\pageref{firstpage}--\pageref{lastpage}} \pubyear{}

\maketitle

\label{firstpage}

\begin{abstract}
A multiwavelength analysis of star formation associated with the extended green object, {\g19} is presented in this paper. With multiple detected radio and millimetre components, {\g19} unveils as harbouring a protocluster rather than a single massive young stellar object. We detect an ionized thermal jet using the upgraded Giant Meterwave Radio Telescope, India, which is found to be associated with a massive, dense and hot {\it ALMA} 2.7\,mm core driving a bipolar CO outflow. Near-infrared spectroscopy with UKIRT-UIST shows the presence of multiple shock-excited {\h2} lines concurrent with the nature of this region. Detailed investigation of the gas kinematics using {\it ALMA} data reveals {\g19} as an active protocluster with high-mass star forming components spanning a wide evolutionary spectrum from hot cores in accretion phase to cores driving multiple outflows to possible UC{\hii} regions.
\end{abstract}


\section{INTRODUCTION}
The last decade has seen a plethora of studies focussed on understanding the formation mechanism of massive ($\rm M \gtrsim 8M{_\odot}$) stars. Recent reviews by \citet{2014prpl.conf..149T} and \citet{2018ARA&A..56...41M} detail the significant progress made in the field. State-of-the-art numerical simulations \citep[e.g.][]{{2007ApJ...656..959K},{2010ApJ...722.1556K}} and tremendous advancement in observational facilities ({\it Spitzer} Space Telescope, {\it Herschel} Space Observatory, {\it ALMA}) have enabled us to probe the intricacies involved and get a better insight of the complex processes associated with high-mass star formation. 

\begin{figure*}
\hspace*{-0.6cm}
\centering 
\includegraphics[scale=0.20]{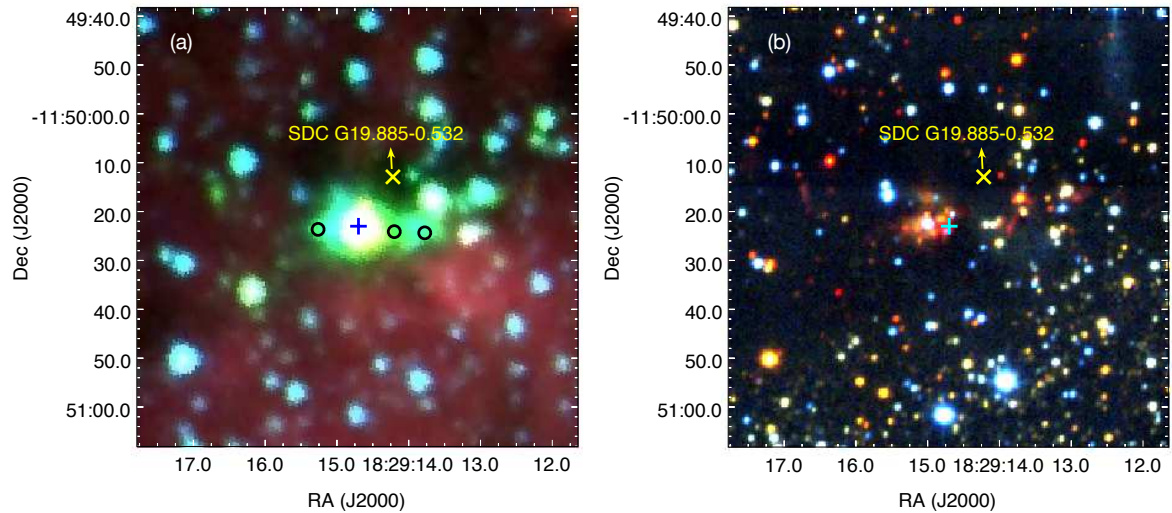}
\caption{(a)IRAC colour composite image of the region around {\g19} where the 3.6, 4.5, and 8.0\,{\um} bands are displayed in red, green, and blue, respectively. Location of the identified IRDC is shown with the `$\times$' symbol. The `$+$' marks the position of the MYSO located at 4.5\,{\um} emission peak and the black circles are the positions of the EGO knots discussed by \citet{2010AJ....140..196D}. (b)UKIDSS colour composite image of the same region. Here, $J$ (1.25\,{\um}), $H$ (1.63\,{\um}) and $K$ (2.20\,{\um}) band data are shown in red, green, and blue, respectively.}
\label{irac_ukidss_rgb}
\end{figure*}

\par Of particular interest are the initial phases of massive star formation, which are observationally the most challenging. The discovery of an interesting population of `extended green objects' (EGOs) \citep{{2008AJ....136.2391C},{2009ApJS..181..360C}}, with the {\it Spitzer} Galactic Legacy Infrared Mid-Plane Survey Extraordinaire (GLIMPSE) \citep{2003PASP..115..953B}, has created a unique database to systematically probe these early phases and aptly constrain the proposed theories. Dedicated studies spanning various frequency regimes have been conducted on EGOs and they are now regarded as signposts of high-mass star formation where the enhanced, extended emission in the IRAC 4.5 $\rm \mu m$ band (colour-coded green in the IRAC colour composite images and hence the nomenclature) is attributed to shocked emission in outflows from massive YSOs (MYSOs).

\par Several studies have been carried out to decipher the population of identified EGOs \citep[e.g.][]{{2004ApJS..154..333M},{2004ApJS..154..352N},{2005ApJ...630L.181R},{2005MNRAS.357.1370S}}, but focussed multiwavelength studies towards individual objects are still few. Such studies are crucial in gaining deeper insights into the nature of these objects. In this paper, we conduct a rigorous multiwavelength investigation of the region associated with the EGO, {\g19}, which is catalogued as a ``likely" outflow candidate associated with the infrared source, IRAS\,18264-1152 \citep{2008AJ....136.2391C}. The kinematic distance ambiguity to {\g19} has been resolved and it is placed at the near distance of 3.31\,kpc \citep{{2009ApJ...699.1153R},{2014MNRAS.445.1170G}}.
The bolometric luminosity of the Red MSX Source (RMS), G019.8817-00.5347, associated with IRAS\,18264-1152 is estimated to be $7.8\times10^3\,L_\odot$ \citep{2013ApJS..208...11L}. This number has been recently refined to $4.7\times10^3\,L_\odot$ in the RMS survey website\footnote{\url{http://rms.leeds.ac.uk/cgi-bin/public/RMS_DATABASE.cgi}}. Class I and II methanol masers and $\rm H_2O$ maser are detected towards the region \citep{{2002ApJ...566..931S},{2002A&A...390..289B},{2011ApJS..196....9C}}. 
{\fig}\ref{irac_ukidss_rgb} displays the infrared (IR) morphology of {\g19}. The colour composite images use the {\it Spitzer}-IRAC \citep{2004ApJS..154...10F} and the UKIDSS-Galactic Plane Survey \citep{2008MNRAS.391..136L} bands. Both plots show extended emission in the east-west direction with the enhanced IRAC 4.5 $\rm \mu m$ emission elongated westwards beyond the UKIDSS $K$ band nebulosity. The {\it Simbad} database also reveals the presence of an infrared dark cloud (IRDC) possibly associated with {\g19} and is marked in the figure.   

The structure of the paper, which involves an in-depth observational study of the region associated with {\g19} at infrared, submillimeter and radio wavelengths, is as follows. New observations carried out by us, and the archival data used are discussed in Section~\ref{obs}; the data reduction procedures are also included in this section. Results obtained from multiwavelength data are highlighted in Section~\ref{results}. In Section~\ref{discussion}, we deliberate on the the observational evidence of the presence of a protocluster and the existence of an ionized jet, and discuss on the kinematics of the identified dust cores. Section~\ref{summary} summarizes the results and discussion.

\section{OBSERVATIONS AND ARCHIVAL DATA} \label{obs}

\subsection{Radio continuum observation}
Low-frequency radio emission associated with {\g19} is studied with the upgraded Giant Meterwave Radio Telescope (uGMRT), located near Pune, India. Observations for {\g19} were carried out in Band~4 ($550 - 850$ MHz) and Band~5 ($1000 - 1450$ MHz). Observation and map details are listed in {\tab}\ref{radio_obs}. 
The data were calibrated incorporating spectral index and spectral curvature across the wide band using the Astronomical Image Processing System (AIPS). Data reduction is carried out along the lines discussed in \citet{2019MNRAS.485.1775I}. We refer the reader to this paper for an elaborate discussion of GMRT configuration and data reduction steps. In case of {\g19}, calibrated and channel averaged data are split into sub-bands of bandwidth $\sim 32$ and $\sim 36$\,MHz in Band~4 and Band~5, respectively. This is done with the envisaged goal of obtaining in-band spectral index of the radio source. We retain six clean sub-bands in Band~4 and three in Band~5, with central frequencies 651.4, 682.5, 714.1, 745.9, 777.1, and 808.3\,MHz and 1320.7, 1355.8, and 1391.6\,MHz, respectively. Continuum maps for each sub-band were generated separately by limiting the {\it uv} coverage to a common range of $\rm 0.15 - 36\, K\lambda$ with beams matched to 11.5\,arcsec$\times$7.6\,arcsec, the largest beam among the six sub-bands. Similarly, for Band~5 the {\it uv} range is fixed to $\rm 0.17 - 100\,K\lambda$ and beams matched to 4.3\,arcsec$\times$2.5\,arcsec. As discussed in \citet{2019MNRAS.485.1775I}, scaling factors are estimated to range between 1.65 to 2.10 and 1.22 to 1.24 within Band~4 and Band~5, respectively. Using these, the primary beam corrected and self-calibrated sub-band maps are scaled to account for variations in the system temperatures of the antennas at the GMRT while observing targets located in the Galactic plane.
.

\begin{table}
\caption{GMRT observations and map details for {\g19}}
\begin{center}
\centering
\begin{tabular}{l l l} \hline \hline \
Details 				 & Band~4		& Band~5 \\
\hline \
Date of Obs.		& 19 June 2018 	& 18 June 2018 \\
Flux calibrators  & 3C286  	&	3C286 \\
Phase calibrators 	& 1822-096 &1743-038 \\
Integration time & $\sim 5$~hrs & $\sim 5$~hrs \\
Synthesized beam &  $\sim$ 11.5\,arcsec  &$\sim$ 4.3\,arcsec \\
					&$\times$ 7.6\,arcsec  &$\times$ 2.7\,arcsec \\
{\it rms} ($\rm \mu Jy\,beam^{-1}$)   & $100-150$ & $44-47$ \\
\hline \
\end{tabular}
\label{radio_obs}
\end{center}
\end{table}

\subsection{Near-infrared spectroscopic observation}
We carried out the near-infrared (NIR) spectroscopic observations towards {\g19} with the 3.8-m United Kingdom Infrared Telescope (UKIRT), Hawaii. Observations were obtained on 2 April 2015 with the UKIRT $\rm 1-5\,\mu m$ Imager Spectrometer (UIST, \citealt{2004SPIE.5492.1160R}) using the {\it HK} grism set-up. The slit ($\sim$0.48\,arcsec wide and 120\,arcsec long) was oriented at an angle 88\degree east of north and centered at ($\rm \alpha_{J2000}= 18^{h}29^{m}14.7^s, \delta_{J2000} = -11\degree 50\arcmin 16.7\arcsec$) as shown in {\fig}\ref{nir_image}. The exposure time per frame was 120s and the total on-source integration time was 720s. SAO 142372, an A5 IV/V type star, was also observed, with the same instrument settings as for the target, for telluric and instrumental corrections. Following the standard procedures explained in \citet{2019MNRAS.485.1775I}, UKIRT pipeline ORAC-DR and Starlink packages FIGARO and KAPPA were used for data reduction, spectra extraction and wavelength calibration. As the sky conditions during our observations were not photometric, the final spectra are not flux calibrated.


\subsection{Far-infrared data from Hi-GAL survey}
To understand the physical properties of the cold dust emission associated with {\g19}, we use {\it Herschel} far-infrared (FIR) archival data. The {\it Herschel Space Observatory} houses a 3.5-m telescope with instruments covering the spectral range of $\rm 55-671\,\mu m$ \citep{2010A&A...518L...1P}. 
For our study, we have utilized observations carried out with the Photodetector Array Camera and Spectrometer (PACS, \citealt{2010A&A...518L...2P}) and Spectral and Photometric Imaging Receiver (SPIRE, \citealt{2010A&A...518L...3G}). The $\rm 70-500\,\mu m$ images retrieved were observed as a part of the Herschel infrared Galactic plane Survey (Hi-GAL, \citealt{2010PASP..122..314M}). The images have resolutions of 5, 13, 18.1, 24.9 and 36.4 {\arc} at 70, 160, 250, 300, and 500~{\um}, respectively.

\subsection{APEX+Planck data}
APEX+Planck database provides 870\,{\um} images. These are generated by combining 870\,{\um} LABOCA bolometer array data from the APEX Telescope Large Area Survey of the Galaxy (ATLASGAL, \citealt{2009A&A...504..415S}) and 850\,{\um} data from Planck/HFI. The combined data-set offers improved spatial dynamic range of Galactic cold dust emission \citep{2016A&A...585A.104C}. The retrieved APEX+Planck image of the region associated with {\g19} has pixel size of 3.4{\arc} and angular resolution of 21{\arc}. 

\subsection{Atacama Large Millimeter Array archival data}
{\it Atacama Large Millimeter Array} ({\it ALMA}) archival data at
2.7\,mm (Band~3) and 870\,{\um} (Band~7) are used to study the {\g19} complex. Observations were made during 2017-2018 (PI: S. Leurini \#2017.1.00377.S). The high- and medium-resolution {\it ALMA} observations were obtained with the 12-m array in the FDM Spectral Mode and the low-resolution observations used the 7-m array in the ACA Spectral Mode. The high resolution 2.7\,mm continuum map, with a beam size of 0.46\,arcsec$\times$0.28\,arcsec, is used to identify the compact dust cores associated with {\g19}. Along with the continuum observations, data cubes in both the bands are also retrieved. Three data sets with resolutions 16.9\,arcsec$\times$8.0\,arcsec, 2.6\,arcsec$\times$2.2\,arcsec, and 0.46\,arcsec$\times$0.28\,arcsec and  velocity resolution of $\rm 0.67\,km\,s^{-1}$ in Band~3 ($\rm 84-116\,GHz$) and two data sets with resolutions 5.2\,arcsec$\times$2.6\,arcsec and 0.67\,arcsec$\times$0.47\,arcsec and velocity resolution of $\rm 0.86\,km\,s^{-1}$ in Band~7 ($\rm 275-373\,GHz$) are used. 
In Band~3, {\it rms} of 4.2, 3.6 and 17\,mJy are achieved for high, medium and low resolution observations, respectively. 
In Band~7, {\it rms} of 3.5 and 17\,mJy are achieved for high and low resolution observations, respectively. Analysis of molecular line data is carried out using CLASS90 (Continuum Line Analysis Single-dish Software), a {\small GILDAS}\footnote{\url{http://www.iram.fr/IRAMFR/GILDAS}} software (Grenoble Image and Line Data Analysis  Software).

\section{RESULTS} \label{results}

\subsection{Low-frequency radio emission} \label{radio_text}

\begin{figure*}
\centering 
\includegraphics[scale=0.27]{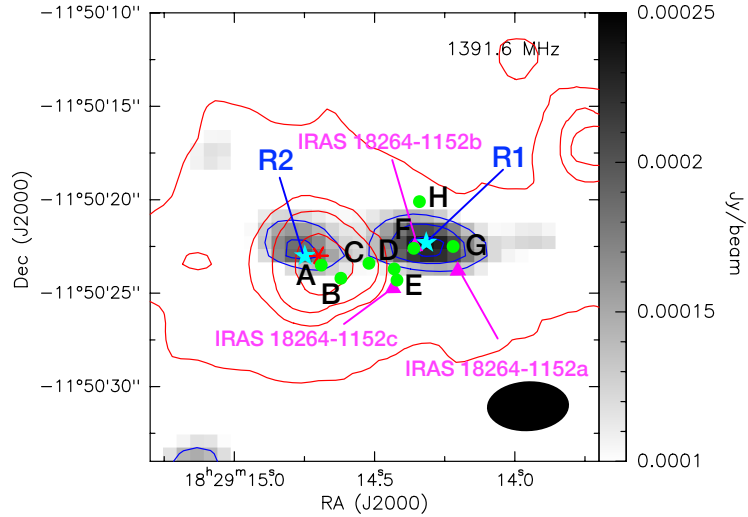} 
\caption{The radio continuum map of {\g19} at 1391.6\,MHz is shown in grey scale with contour levels at 3, 4, 5, and 6$\sigma$ ($\rm \sigma=45\,\mu Jy\,beam^{-1}$). This map has a beam size of 4.3\,arcsec$\times$2.7\,arcsec that is represented by the black ellipse. The positions of the two radio components, R1 and R2 are marked on the image by `$\star$'s. Radio sources $\rm A-H$, identified by \citet{2016ApJS..227...25R}, are marked by filled circles and the three radio components IRAS\,18264-1152a, IRAS\,18264-1152b and IRAS\,18264-1152c, identified by \citet{2006AJ....131..939Z}, are indicated by filled triangles. Contours of the 4.5\,{\um} emission are overlaid in red with the contour levels at 3, 9, 30 and 120$\sigma$ ($\rm \sigma = 5.0\,MJy\,sr^{-1}$). The position of the MYSO \citep{2010AJ....140..196D} associated with {\g19} is indicated with `$\ast$'.}
\label{radio_image}
\end{figure*}

\begin{table}
\caption{Physical parameters of radio components R1 and R2 associated with {\g19}.}
\begin{center}

\hspace*{-1.1cm}
\centering  
\begin{tabular}{c c c c c c c} \hline \hline 
Component & \multicolumn{2}{c}{Peak position}	& Peak flux density  \\
& $\rm \alpha(J2000)\,({^h}\,{^m}\,{^s})$		& $\rm \delta(J2000)\,(\degree\,\arcmin\,\arcsec)$ &($\rm mJy\,beam^{-1}$) \\
\hline \
R1 	& 18 29 14.3 	& -11 50 22.3 	& 0.25$\pm$0.01  \\
R2 	& 18 29 14.7 	& -11 50 23.0 	& 0.20$\pm$0.01	\\
\hline \\	
\end{tabular}
\label{radio_par}
\end{center}
\end{table}

Using the uGMRT, we probe the low-frequency domain of the ionized emission associated with {\g19}. In Band~4, none of the sub-band images have detection above the $3\sigma$ level while faint, elongated emission is seen in the Band~5 maps. The deep uGMRT maps in Band~5 with a {\it rms} level of $\rm \sim 45\,\mu Jy\,beam^{-1}$ enable the detection of weak ionized emission. In comparison, there is no emission detected in the VLA 20~cm map from MAGPIS. The sub-band image at 1391.6~MHz is shown in  {\fig}\ref{radio_image}. This map clearly reveals the presence of an elongated, two-component emission feature in the east-west direction. The two distinct and compact radio components are labelled R1 and R2. The contours of the 4.5\,{\um} emission plotted show that the radio emission is confined to the central part of the EGO. 

The radio emission associated with {\g19} has been studied by \citet{2006AJ....131..939Z}, \citet{2016ApJS..227...25R}, and \citet{2019ApJ...880...99R} using VLA. Based on observations at 3.6\,cm (8.5\,GHz), 1.3\,cm (22.5\,GHz) and 7\,mm (43.3\,GHz), \citet{2006AJ....131..939Z} classify this source as a `triple stellar system' and identify three radio components (IRAS\,18264-1152a, b, and c), the positions of which are shown in {\fig}\ref{radio_image}. In the recent paper, \citet{2016ApJS..227...25R} present high-sensitivity, sub-arcsecond resolution observations at 6\,cm (4.9 and 7.4\,GHz) and 1.3\,cm (20.9 and 25.5\,GHz), where twelve compact radio sources are detected towards {\g19}. Eight among them, sources $\rm A - H$, lie within the $3\sigma$ level of the 4.5\,{\um} emission and are marked {\fig}\ref{radio_image}. The component R1 is co-spatial (within $\sim 1$\,arcsec) with the radio sources, F and IRAS\,18264-1152b. Component R2 has similar positional match with the radio source, A. The position of the MYSO located at the 4.5\,{\um} emission peak \citep{2010AJ....140..196D} also agrees well with the coordinates of R2. However, this component is not detected in the VLA maps presented by \citet{2006AJ....131..939Z}.  

The position of the components and the peak flux densities are listed in {\tab}\ref{radio_par}. Component R1 is barely resolved and component R2 is unresolved in the uGMRT maps. It is thus difficult to get a reliable estimate of the integrated flux density for these unresolved sources with weak emission \citep{{2009A&A...501..539U},{2016ApJS..227...25R}}. Hence, in the estimation of the spectral index we use the peak flux density. Following the discussion in \citet{2009A&A...501..539U}, the upper limits to the deconvolved sizes of R1 and R2 are taken to be half the FWHM of the restored beam size. This gives a value of 2.1\,arcsec$\times$1.3\,arcsec for the components.

 \subsection{Near-infrared line emission} \label{nir_spec}
 
\begin{figure}
\centering 
\includegraphics[scale=0.35]{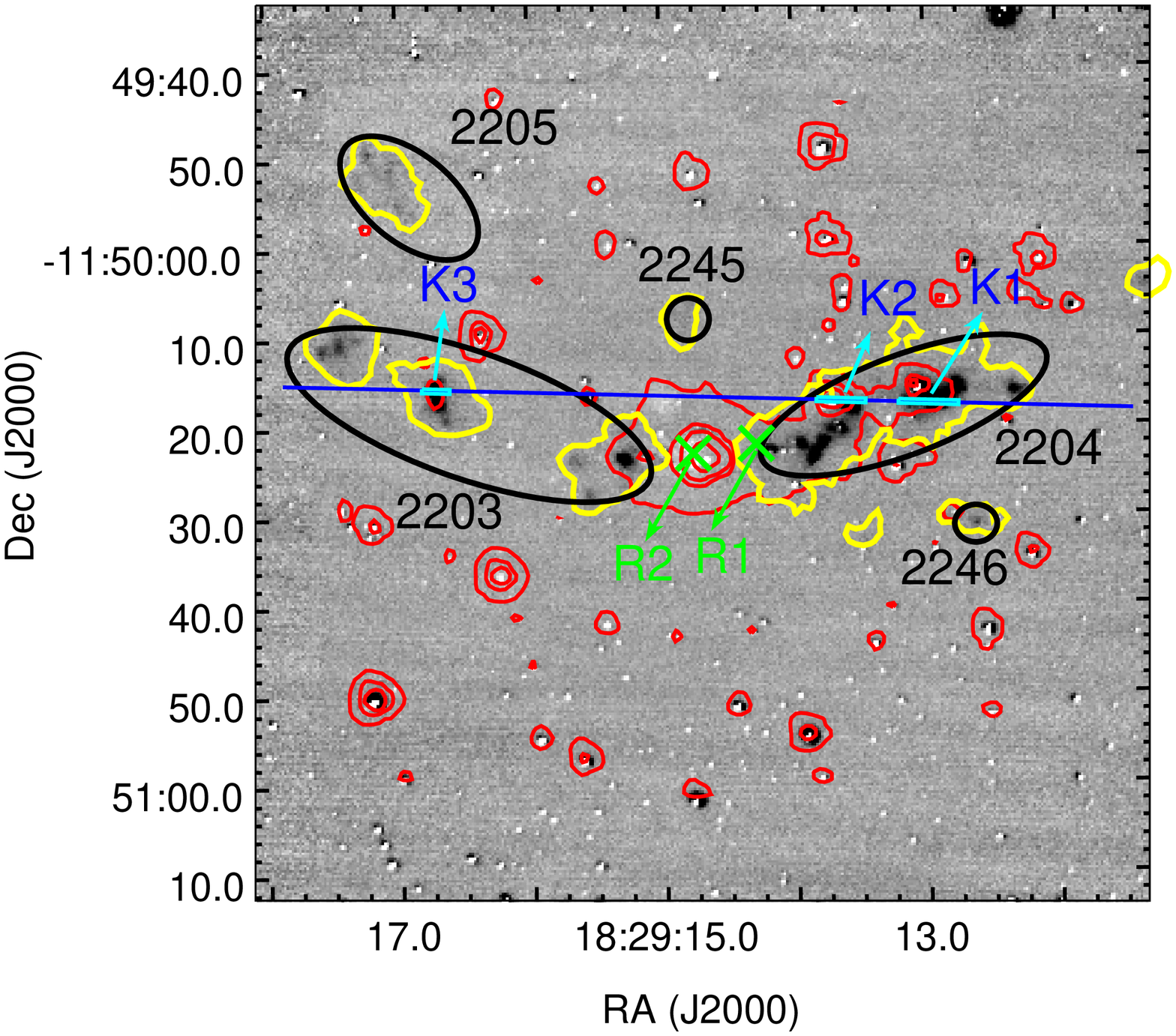}
\caption{Continuum subtracted {\h2} image of {\g19} from \citet{2010MNRAS.404..661V}. The positions of the identified radio components, R1 and R2 are marked by `$\times$'s. The contours of the 4.5\,{\um} emission are given in red with the contour levels same as {\fig}\ref{radio_image}. The blue line indicates the orientation of the slit and the cyan rectangles are the apertures over which the spectra are extracted. The MHOs (2203-2205,2245,2246) identified by \citet{2010MNRAS.404..661V}, \citet{2012ApJS..200....2L}, and \citet{2012MNRAS.421.3257I} are highlighted in the black ellipses. The yellow contours trace the location and extent of the {\h2} knots identified by \citet{2011MNRAS.413..480F}.}
\label{nir_image}
\end{figure}

\begin{figure*}
\centering 
\includegraphics[scale=0.33]{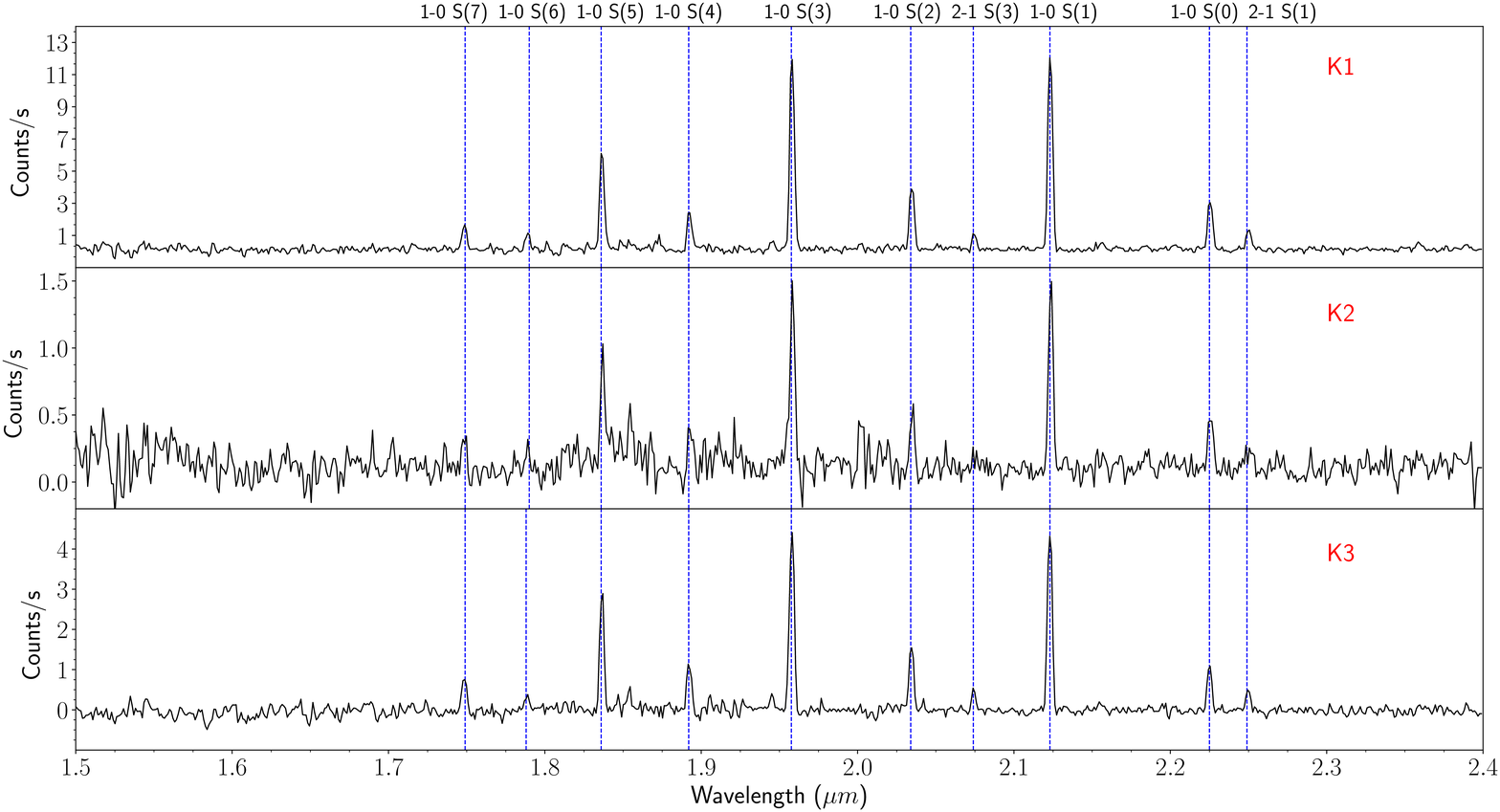}
\caption{The {\it HK} spectrum of {\g19} extracted over the apertures K1, K2 and K3 covering three {\h2} knots  (refer {\fig}\ref{nir_image}). The spectral lines identified along all the three apertures are marked over the spectra and the details of these lines are given in {\tab}\ref{nir_lines}.}
\label{nir_spectrum}
\end{figure*}

As propounded by several authors \citep[and references therein]{2015A&A...573A..82C}, shock-excited lines of {\h2} and {\feii} and broad bandheads of CO are the spectral carriers responsible for the enhanced 4.5\,{\um} emission of EGOs. Of the few spectroscopic studies of EGOs \citep[e.g.][]{{2010AJ....140..196D},{2015A&A...573A..82C},{2016ApJ...829..106O},{2019MNRAS.485.1775I}}, \citet{2010AJ....140..196D} focussed on MIR spectroscopy of two EGOs including {\g19} using the NIRI instrument on the Gemini North telescope. The spectra extracted towards the three associated knots of {\g19}
show the presence of the shock-excited $\rm 0-0~S(9)$ line of {\h2} at 4.695\,{\um}. Moreover no continuum emission was detected towards these knots. In our study, we investigate the NIR regime to identify shock indicators and confirm the association of {\g19} with protostellar outflows.

\par The {\h2} line image towards {\g19}, presented in \citet{2010MNRAS.404..661V} reveals the presence of a bipolar outflow in the east-west direction which is consistent with CO outflow detected by \citet{2002A&A...383..892B}. 
Same results were obtained by \citet{2012ApJS..200....2L} using the UWISH2 survey and they enlist {\g19} as an EGO with a large-scale (angular scale of $\sim$~1.6\,arcmin) bipolar outflow. In {\fig}\ref{nir_image} we display the continuum-subtracted {\h2} line image from \citet{2010MNRAS.404..661V} where the orientation of the slit position for UIST observations is marked. The identified apertures, K1, K2, and K3, over which the spectra are extracted are also highlighted in the figure.  
Molecular hydrogen objects (MHOs) from the catalogue of molecular hydrogen emission-line objects\footnote{\url{http://cdsweb.u-strasbg.fr/cgi-bin/qcat?J/A+A/511/A24}} \citep{2010A&A...511A..24D} are also indicated. MHO~2245 and MHO~2246 are new additions to the catalogue by \citet{2012ApJS..200....2L} and \citet{2012MNRAS.421.3257I}, respectively. Identified locations and extent of {\h2} knots from \citet{2011MNRAS.413..480F} are also shown in the figure. Apertures K1 and K2 probe the knots in MHO~2204 and K3 samples the central knot in MHO~2203. The extracted spectra are shown in {\fig}\ref{nir_spectrum} where we see strong detections of several {\h2} lines. The lines detected and their corresponding wavelengths are listed in {\tab}\ref{nir_lines}.  

\begin{table}
\caption{Lines detected in the spectra extracted along the apertures K1, K2 and K3 towards {\g19}.}
\begin{center}
\begin{tabular}{c c} \hline \hline \
Line		& Wavelength ({\um}) 	\\
\hline \
{\h2} $1-0$ S(7) & 1.7480 \\
{\h2} $1-0$ S(6) & 1.7880\\
{\h2} $1-0$ S(5) & 1.8358\\
{\h2} $1-0$ S(4) & 1.8920\\
{\h2} $1-0$ S(3) & 1.9576\\
{\h2} $1-0$ S(2) & 2.0338\\
{\h2} $2-1$ S(3) & 2.0735\\
{\h2} $1-0$ S(1) & 2.1218\\
{\h2} $1-0$ S(0) & 2.2235\\
{\h2} $2-1$ S(1) & 2.2477\\
\hline \
\end{tabular}
\label{nir_lines}
\end{center}
\end{table}

\par The {\h2} lines detected in the spectra could have either a thermal or a non-thermal origin. Jets and outflows that are heated to typically few 1000\,K can give rise to thermal emission from shock fronts. Whereas, non-thermal emission is ascribed to fluorescence produced by non-ionizing UV photons. Shock excited emission arises from low levels of excitation, whereas UV fluorescence populates both high and low-{\it v} states \citep{{2003MNRAS.344..262D},{2015A&A...573A..82C}}. Our spectra do not show the presence of transitions with {\it v}\,$\geq$\,6. In addition, the ratio of $\rm 1-0\,S(1)/2-1\,S(1)$ is estimated to be $\sim 8:1$. These suggest the origin of the detected {\h2} lines to be due to thermal excitation.
However, the absence of relevant fluorescent {\h2} emission in {\g19} can also be attributed to high extinction that screens the UV photons emanating from the central star. Nonetheless, morphological resemblance of the extended {\h2} emission knots to a bipolar jet and the association with an outflow source supports the shock excitation scenario. The shock-excited origin of the detected $\rm H_2$ lines confirms that the transitions of this molecule within the IRAC 4.5\,{\um} band are among the spectral carriers responsible for the enhanced `green' emission that classifies {\g19} as an EGO.

\subsection{Emission from dust} \label{dust}
\begin{figure*}
\centering
\includegraphics[scale=0.20]{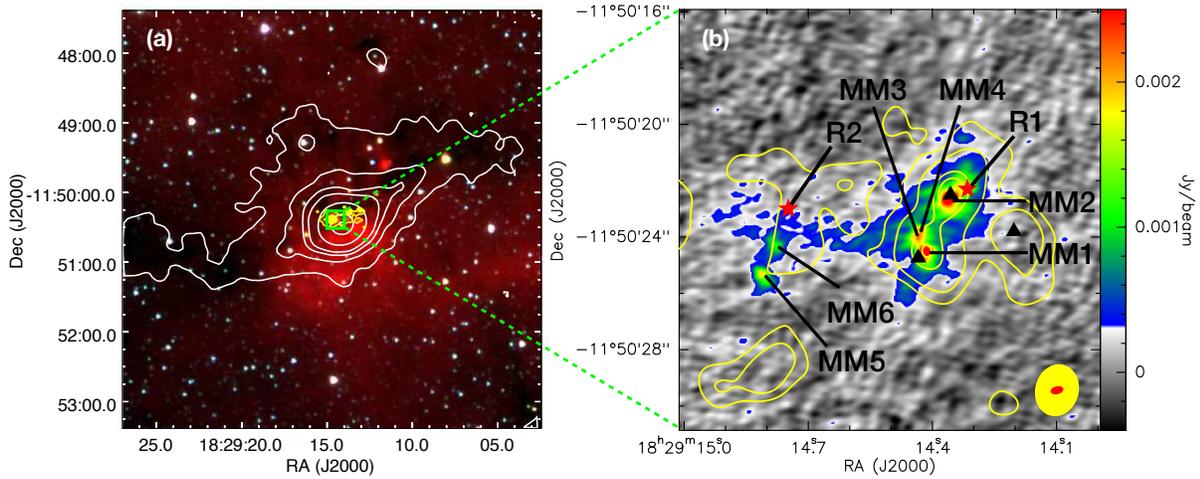}
\caption {(a) Colour composite image of the region associated with {\g19}, using IRAC 3.6\,{\um} (blue), 4.5\,{\um} (green) and 8.0\,{\um} (red) bands. The white contours depict the APEX+Planck 870\,{\um} emission with levels 2, 3, 4, 5, 9, 20, and 26$\sigma$ ($\rm \sigma=0.3\,Jy\,beam^{-1}$). The yellow contours represent the 4.5\,{\um} emission with the contour levels same as in {\fig}\ref{radio_image}. (b) The high resolution 2.7\,mm {\it ALMA} map towards {\g19} is depicted in the colour scale. The positions of the two radio components and the six mm peaks identified are marked on the map. The yellow contours represent emission at 7\,mm from \citet{2006AJ....131..939Z} with contour levels at 4, 8, 18, 22, and 26$\sigma$ ($\rm \sigma=53.8\,{\mu}Jy\,beam^{-1}$). The filled black triangles are the positions of the radio peaks identified by these authors. The beams of the 2.7 and 7\,mm maps are shown as filled red and yellow ellipses, respectively, towards the bottom right of the image.}
\label{870_alma_cores}
\end{figure*}

IRAC bands trace the thermal emission associated with the warm dust component and emission from the polycyclic aromatic hydrocarbons excited by the UV photons in the photodissociation regions. However, emission from the stellar photosphere dominates the shorter wavelength bands (3.6 and 4.5\,{um})  \citep{2008ApJ...681.1341W}.
In addition to this, the 4.5\,{\um} band includes emission from shock-excited {\h2} lines and CO ({\it v$=1-0$}) band at 4.6\,{\um} with contribution from the Br$\alpha$ and Pf$\beta$ lines.
 The spatial distribution of dust emission towards {\g19} is shown in {\fig}\ref{870_alma_cores}(a), which is a colour composite image generated using the IRAC 8.0, 4.5, and 3.6\,{\um} bands. The APEX+Planck 870\,{\um} contours are overlaid along with the contours of the 4.5\,{\um} emission. A clump is clearly identified at 870\,{\um}. Warm dust sampled at 8.0\,{\um} is seen to be extended towards the south and west of the clump and, at 870\,{\um}, a clearly discernible large-scale filamentary structure is visible with the clump located within it.   

\subsubsection{Dust clump} \label{dust_clump}

\begin{figure}
\centering
\includegraphics[scale=0.42]{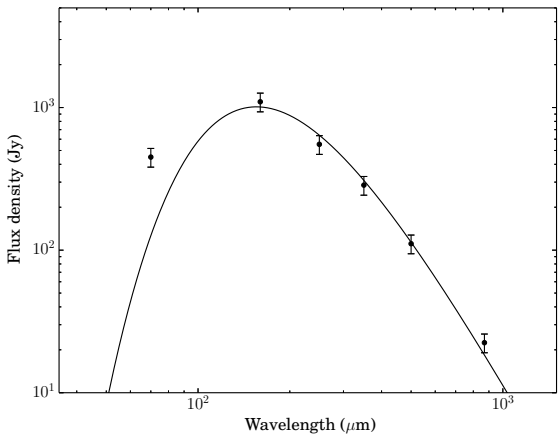}
\caption {Spectral energy distribution of the APEX+Planck clump associated with {\g19} in the wavelength range 70 to 870\,{\um}. The integrated flux density within the clump, represented by black circles is plotted in log-scale with a 15\% error bar. The best-fit modified blackbody model is represented by the solid curve. The data point corresponding to 70\,{\um} is excluded in the SED fitting.}
\label{sed}
\end{figure}


From the APEX+Planck 870\,{\um} map, a clump is identified using the two-dimensional {\it Clumpfind} procedure \citep{1994ApJ...428..693W} with a 3$\sigma$ ($\rm \sigma = 0.3\,Jy\,beam^{-1}$) threshold and optimum contour levels. The aperture of this clump coincides with the 3$\sigma$ level contour of the 870\,{\um} emission.
To investigate the nature of the cold dust associated with the identified clump, we assume the FIR emission to be optically thin and model the same with a modified single-temperature blackbody following the formalism discussed in detail in \citet{2019MNRAS.485.1775I}. The SED and the best fit modified blackbody are shown in {\fig}\ref{sed}. The model derived dust temperature and line-of-sight column density values are 18.6$\pm$1.2\,K	and  $5.3\pm1.0\times 10^{22}\,\rm cm^{-2}$, respectively. The mass of the clump is estimated to be 1911\,$M_\odot$. This is in good agreement with the estimate of $\sim 2100\,M_\odot$ derived by \citet{{2002ApJ...566..945B},{2005ApJ...633..535B}} from the 1.2\,mm emission. Both these values derived from dust emission are lower by a factor of $\sim 2$ in comparison with the value of $3900\,M_\odot$ obtained from the $\rm H^{13}CO^+$ line emission \citep{2007ApJ...654..361Q}. The discrepancy could be due to the fact that the $\rm H^{13}CO^+$ line traces the lower density gas and could even be slightly optically thick as opposed to the optically thin assumption we have employed while estimating the clump mass. Moreover, the $\rm H^{13}CO^+$ is likely to be affected by outflows and shocks associated with the clump. The derived clump parameters are listed in Table \ref{clump_properties}.

\begin{table*}
\caption{Physical parameters derived for the clump identified to be associated with {\g19}. The peak position of the 870\,{\um} emission, radius, mean temperature and column density, mass, and volume number density of the clump are listed.}
\begin{center}

\hspace*{-1.1cm}
\centering  
\begin{tabular}{c c c c c c c} \hline \hline 
\multicolumn{2}{c}{Peak position} 			&Radius		&Mean $T_d$		&Mean $N(\rm H_2)$		& Mass			& Number density, $n(\rm H_2)$		\\
$\rm \alpha(J2000)\,({^h}\,{^m}\,{^s})$		& $\rm \delta(J2000)\,(\degree\,\arcmin\,\arcsec)$		&(pc)		&(K)			& ($\rm 10^{22}\,cm^{-2}$)			& ($M_\odot$)		&($\rm 10^{3}\,cm^{-3}$) 	\\
\hline \
18 29 14.14		& -11 50 28.57 		& 0.7	&	18.6$\pm$1.2		& 5.3$\pm$1.0		&1911		& 18.0	\\
\hline \\	
\end{tabular}
\label{clump_properties}
\end{center}
\end{table*}

\subsubsection{Millimeter cores} \label{mm_cores}

\begin{table*}
\caption{Physical parameters of the 2.7\,mm cores associated with {\g19}.}
\begin{tabular}{c c c c c c c c c } \hline \hline 

Core 	 &\multicolumn{2}{c}{Peak position} & Deconvolved size  &Integrated flux density	& Peak flux density   & $ V_{\rm LSR} $  & $\Delta V$ \\
			 &RA (J2000) $({^h}~{^m}~{^s})$			&Dec (J2000)	$(\degree~\arcmin~\arcsec)$	 & (arcsec$\times$arcsec)			& (mJy)			& (mJy/beam) 	&($\rm km\,s^{-1}$)  &($\rm km\,s^{-1}$) \\			
\hline \\
MM1		&18 29 14.4 	&-11 50 24.5 		& 0.65$ \times $0.44	&10.7		&3.4		& 44.2		& 6.0 \\	
MM2 	&18 29 14.4		&-11 50 22.6		&1.49$ \times $0.94		&30.9		&3.5  		& 45.3	 	& 5.7 \\
MM3		&18 29 14.4		&-11 50 24.0		&0.98$ \times $0.50		&11.9$^\ast$	&2.3$^\ast$	& 43.3$^\ast$	& 4.5$^\ast$ \\
MM4		&18 29 14.4		&-11 50 23.7		&		0.98$ \times $0.50				& 	11.9$^\ast$	& 	2.3$^\ast$	& 43.3$^\ast$ 	& 	4.5$^\ast$		\\
MM5		&18 29 14.8		&-11 50 25.3		&0.74$ \times $0.26		&5.4			&1.9		& 43.3	 	& 2.3 \\
MM6		&18 29 14.8		&-11 50 24.3		&0.62$ \times $0.46		&3.9			&1.1		& 42.3	 	& 1.6\\
\hline \\
\end{tabular}
\label{mm_table}

$^\ast$ MM3 and MM4 are unresolved in the map, hence the quoted deconvolved size, integrated, peak flux density, LSR velocity, and velocity width refer to the combined region covering both the cores. 
\end{table*}

The high-resolution {\it ALMA} 2.7\,mm dust continuum map displayed in {\fig}\ref{870_alma_cores}(b) shows the presence of six dense, compact dust cores labelled $\rm MM1 - MM6$. Contours of the 7\,mm VLA map from the study by \citet{2006AJ....131..939Z} is overlaid on the image. The 7\,mm emission is more extended, especially in the east-west direction. In an earlier study, \citet{2007ApJ...654..361Q} have mapped IRAS\,18264-1152 at 1.3 and 3.4\,mm with Plateau de Bure Interferometer. They identify two peaks at both the wavelengths. Their western peak seems to be in the vicinity of cores, $\rm MM1 - MM4$ and the eastern component likely associated with cores, MM5 and MM6. Parameters of the identified mm cores are listed in Table \ref{mm_table}. 
The deconvolved sizes of the cores are estimated by fitting 2D Gaussians to each component using the 2D fitting tool of {\small CASA} viewer. MM3 and MM4 are unresolved in the map, hence the quoted size, integrated and peak flux density refer to the combined region covering both the cores.

\subsection{Molecular line emission} \label{molecular}
Molecular line observations are tools for understanding the kinematics and chemical environment of a molecular cloud and and its stage in the evolutionary sequence. In this work, we use archival data from {\it ALMA} to study the emission from molecular lines in the region associated with {\g19}.  

\subsubsection{$\rm CH_3OH$ line emission}
\begin{figure}
\centering 
\includegraphics[scale=0.44]{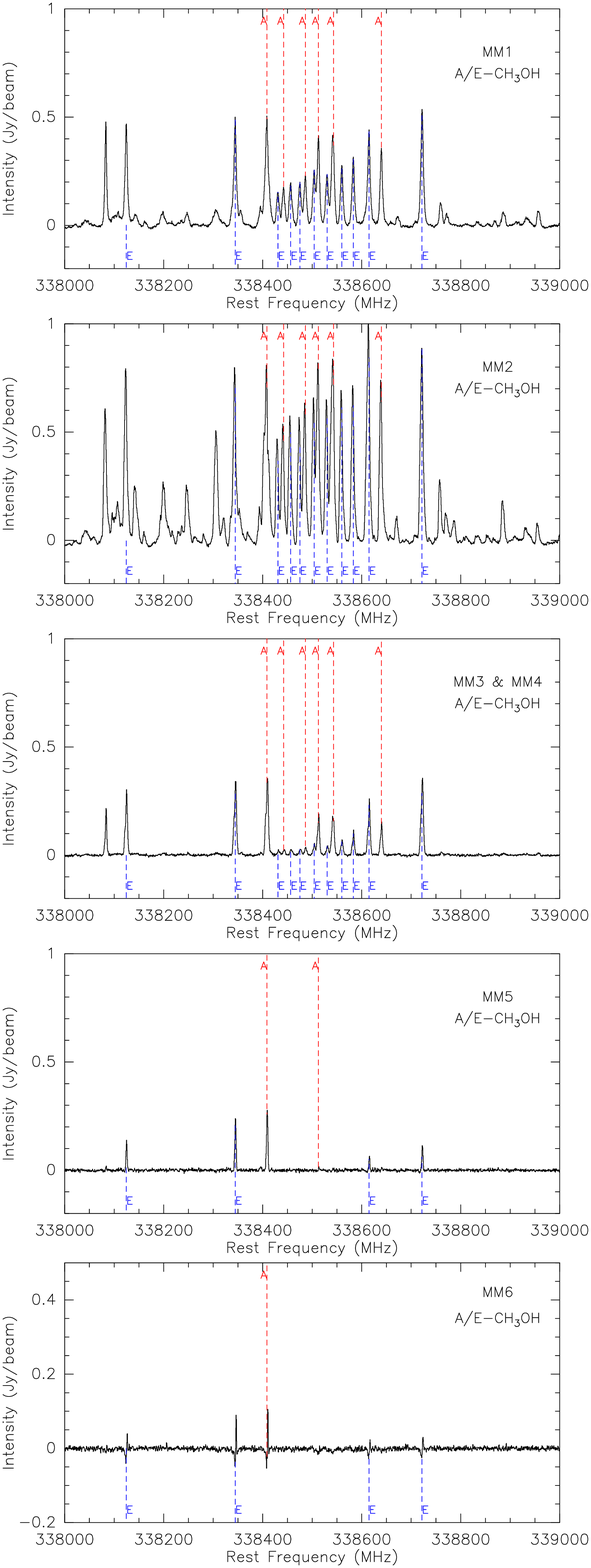}
\caption{Spectra of the $A/E-\rm CH_3OH$ lines detected towards the mm cores associated with {\g19} extracted from the high-resolution Band~7 spectral cube. The dashed red and blue lines indicate the peak positions of the $A-\rm CH_3OH$ and $E-\rm CH_3OH$ lines, respectively.}
\label{ch3oh}
\end{figure}

A forest of $\rm CH_3OH$ lines are detected towards the mm cores in {\it ALMA} Band~7 $\rm (275-373\,GHz)$ within a frequency interval of 3\,GHz $\rm (337.1-340.1\,GHz)$. This is expected as the $\rm CH_3OH$ molecule is believed to be present in massive star forming regions from the very early stages \citep{2011A&A...533A..24W}. For identification of the lines, the data cube is corrected for the LSR velocity of $\rm 43.6\,km\,s^{-1}$ \citep{{2007ApJ...654..361Q},{2002ApJ...566..945B}}. Subsequently, the detected lines are identified by comparing with the rest frequencies reported in the molecular databases, NIST\footnote{\url{http://physics.nist.gov/cgi-bin/micro/table5/start.pl}} \citep{2004JPCRD..33..177L}, CDMS\footnote{\url{https://cdms.astro.uni-koeln.de/cdms/portal}} \citep{2005JMoSt.742..215M}, and JPL\footnote{\url{https://spec.jpl.nasa.gov}} \citep{1998JQSRT..60..883P}. The CDMS and JPL databases are interactively loaded on the spectra employing the Weeds extension of CLASS90. The extracted spectra towards the mm cores are plotted in {\fig}\ref{ch3oh} displaying the identified set of $A-$ and $E-\rm CH_3OH$ lines corresponding to the $J=7-6$ transition. It should be noted here that for cores MM3 and MM4, the separation is less than the beam size, hence the combined spectrum is plotted in the figure. $\rm CH_3OH$ lines are seen to be much stronger towards MM1 and MM2 as compared to the other cores. The accurate identification of the detected lines enables us to determine the LSR velocities of the individual cores which are listed in {\tab}\ref{mm_table}.  

\subsubsection{Emission in isotopologues of CO}
\begin{figure*}
\centering 
\includegraphics[scale=0.30]{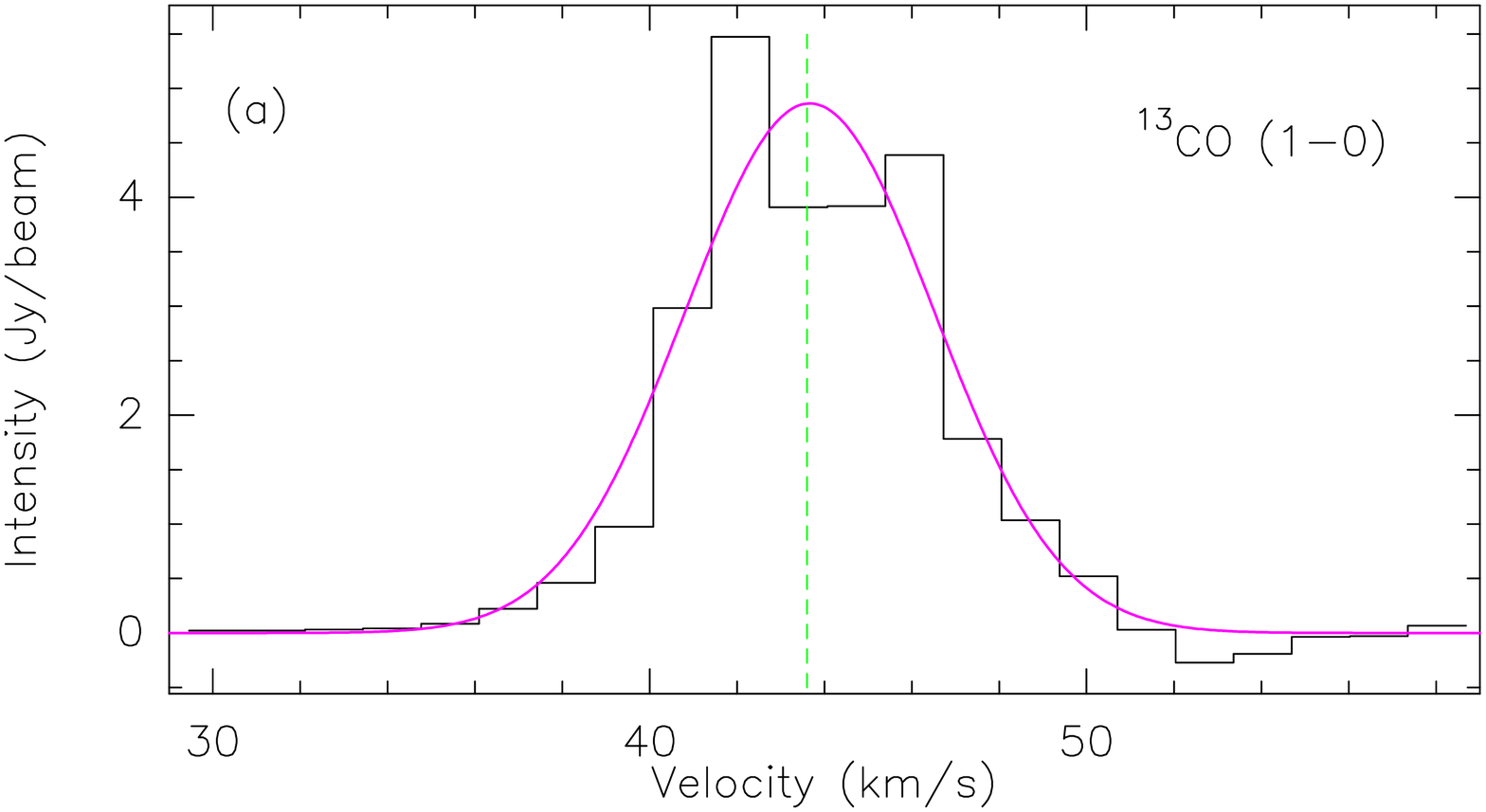} \quad\includegraphics[scale=0.30]{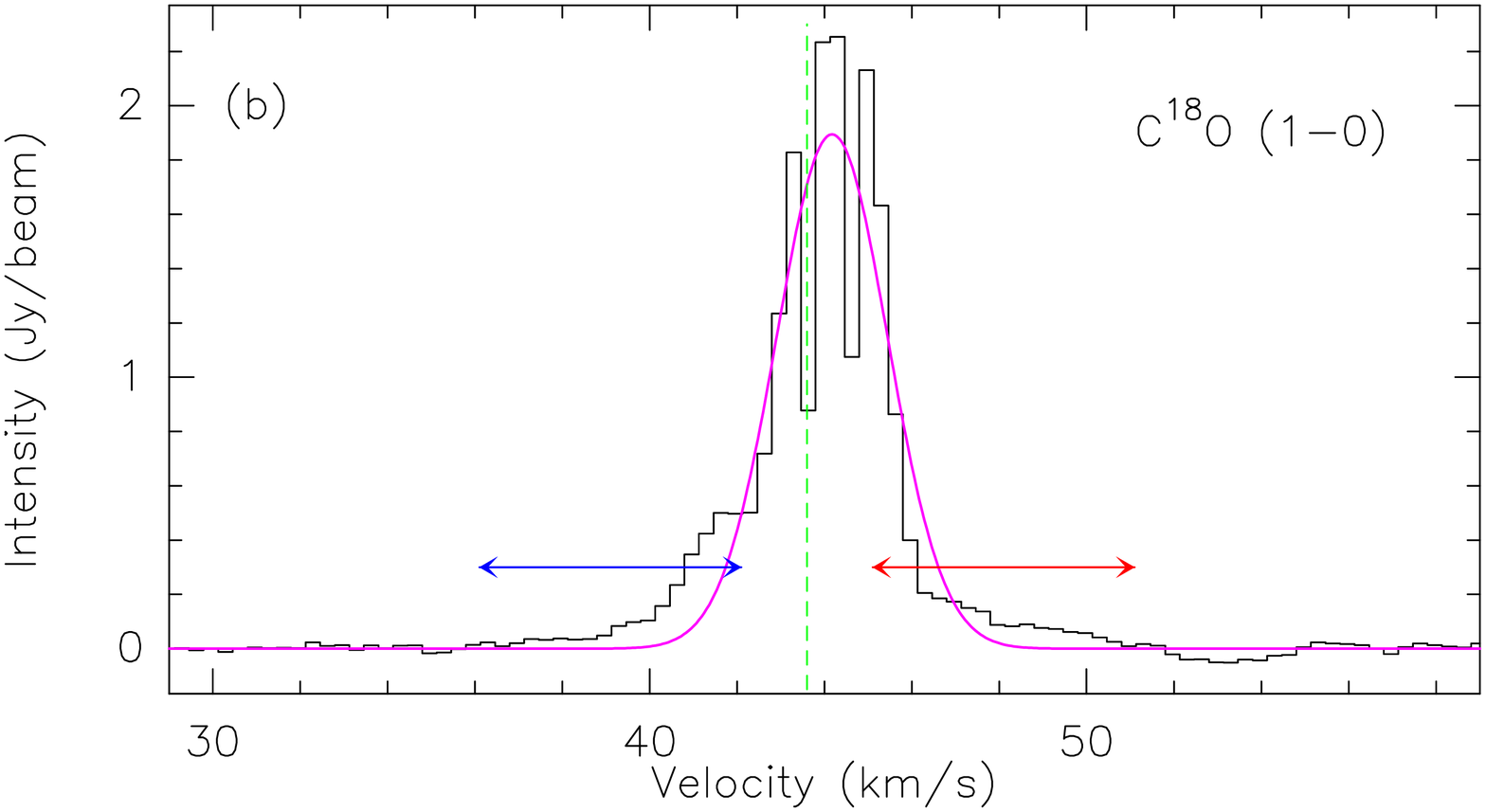}
\caption{Low-resolution spectra of the $\rm ^{13}CO\,(1-0)$ and $\rm C^{18}O\,(1-0)$ lines towards {\g19}. The area over which the spectra are extracted covers all the cores. The vertical green line denotes the LSR velocity, $\rm 43.6\,km\,s^{-1}$. The red and blue arrows in (b) spans the range over which the integrated intensity map is constructed to trace the outflow wings (discussed in Section~\ref{outflow}).}
\label{co_low}
\end{figure*}
\begin{figure*}
\centering 
\includegraphics[scale=0.42]{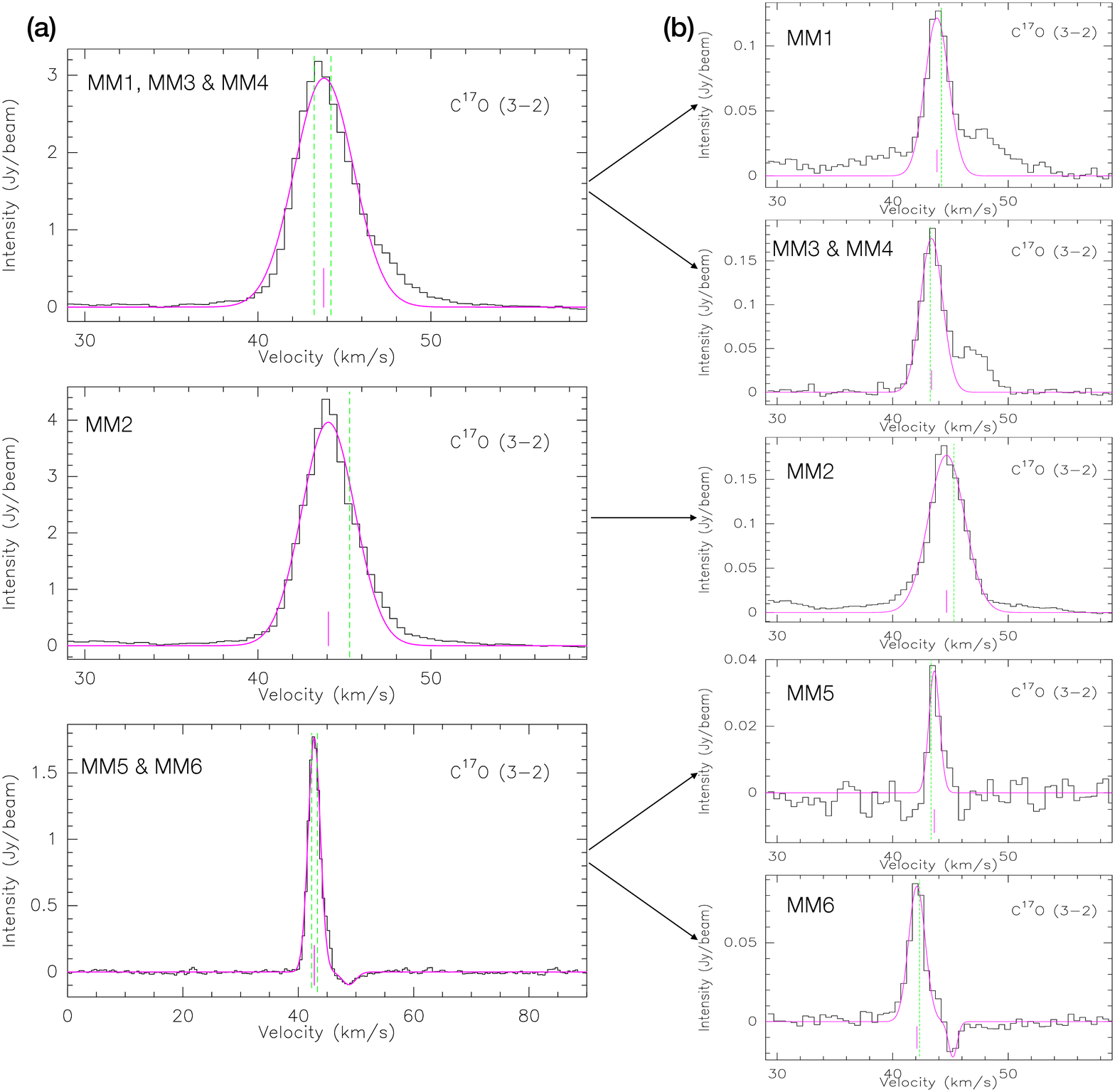}
\caption{The $J=3-2$ transition of $\rm C^{17}O$ towards the mm cores associated with {\g19} from the (a) low-resolution (5.2\,arcsec$\times$2.6\,arcsec) data averaged over the beam and (b) high resolution (0.67\,arcsec$\times$0.47\,arcsec) data averaged over a region covering each core. The magenta curve represents the best fit to each spectrum. Double Gaussians are used to fit the spectra towards MM5 and MM6 (low-resolution) and MM6 (high-resolution). The dashed green line denotes the LSR velocity of each core. The magenta lines in the spectra mark the emission peaks.}
\label{c17o}
\end{figure*}

\begin{figure*}
\centering 
\includegraphics[scale=0.42]{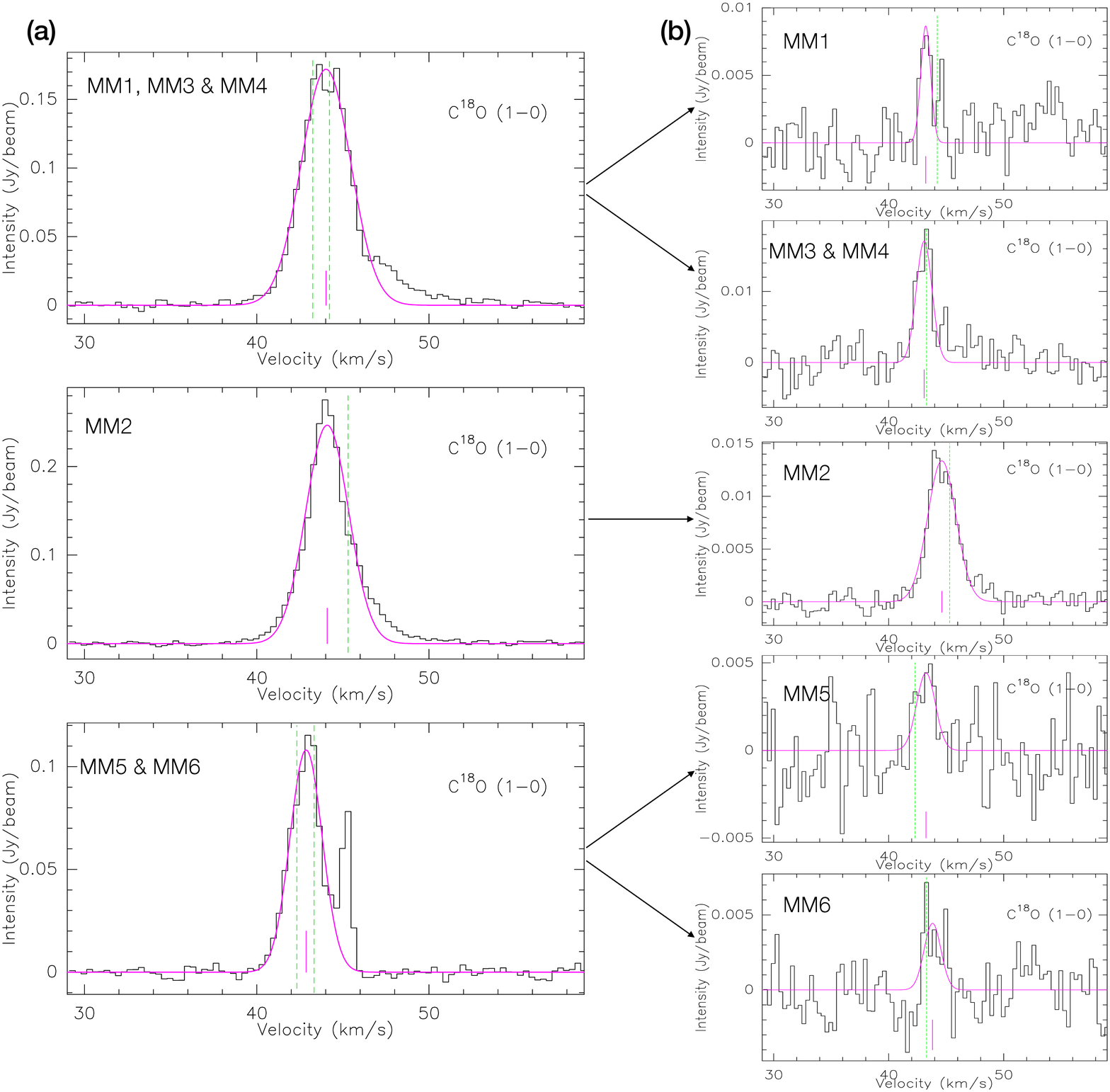}
\caption{Same as {\fig}\ref{c17o} for the $\rm C^{18}O\,(1-0)$ transition detected towards the mm cores associated with {\g19} from the (a) medium-resolution (2.6\,arcsec$\times$2.2\,arcsec) data extracted over the same region as in {\fig}\ref{c17o}(a) and (b) high-resolution (0.46\,arcsec$\times$0.28\,arcsec) data averaged over a region covering each core.}
\label{c18o}
\end{figure*}

\begin{table*}
\caption{Parameters of the isotopologues of the CO molecule detected towards {\g19} from spectral data cubes at different spatial resolutions. The LSR velocity ($ V_{\rm LSR} $), line width ($\Delta V $), intensity and velocity integrated intensity ($\int I_\nu dV$) of each transition are obtained from the single or double Gaussian profiles fits to the extracted spectra.}
\begin{center}
\begin{tabular}{l l l l l l } \hline \hline 

Core 	&$ V_{\rm LSR} $	& $\Delta V$ 	& Intensity 	& $\int I_\nu dV$	\\

&($\rm km\,s^{-1}$)		&($\rm km\,s^{-1}$)	&($\rm Jy\,beam^{-1}$)		&($\rm Jy\,beam^{-1}\,km\,s^{-1}$)		 \\

\hline \
$\rm ^{13}CO\,(1-0)$ \\
\\
Low-resolution (16.9\,arcsec$\times$8.0\,arcsec) \\
MM1-MM6		& 43.7	& 6.7		& 4.9 	& 34.8 \\
	
\hline \

$\rm C^{17}O\,(3-2)$ \\
\\
Low-resolution (5.2\,arcsec$\times$2.6\,arcsec) \\
MM1, MM3 \& MM4		& 43.8		& 4.0		& 3.0	& 12.7 \\
MM2								& 44.1		& 3.6		& 4.0	& 15.3 \\
MM5 \& MM6				& 42.8 (B)	& 2.5 (B)	& 1.8 (B)	& 4.6 (B) \\
									& 48.8 (R)	& 2.8 (R)	&-0.1 (R)	&-0.3 (R) \\
\\
High-resolution (0.67\,arcsec$\times$0.47\,arcsec) \\
MM1					& 43.8 	& 2.4		& 0.1	& 0.3  \\
MM2					& 44.7		& 3.7		& 0.2	& 0.7 	\\
MM3 \& MM4	& 43.4		& 2.3 		& 0.2 	& 0.4  \\
MM5 				& 43.6 	& 1.1 		& 0.04	& 0.04  \\
MM6 				& 42.1 (B)	& 1.7 (B)		& 0.09  (B)	& 0.2 (B) \\
						& 45.2 (R)	& 0.9 (R)		& -0.02 (R)	&-0.02 (R) \\

\hline \

$\rm C^{18}O\,(1-0)$ \\
\\
Low-resolution (16.9\,arcsec$\times$8.0\,arcsec) \\
MM1-MM6		&44.2	& 3.0		& 1.9 	& 6.0 \\
\\
Medium-resolution (2.6\,arcsec$\times$2.2\,arcsec) \\
MM1, MM3 \& MM4		& 44.0		& 3.4		& 0.2	& 0.6 \\
MM2								& 44.1		& 2.9		& 0.2	& 0.8 \\
MM5 \& MM6				& 42.9		& 2.2 		& 0.1 	& 0.3  \\

\\

High-resolution (0.46\,arcsec$\times$0.28\,arcsec) \\
MM1			& 43.2 	& 1.0 	& 0.01	& 0.01  \\
MM2			& 44.6		& 2.9	& 0.01	& 0.04 \\
MM3 \& MM4	& 43.1	& 1.6	& 0.02	& 0.03 \\
MM5 		& 43.8		& 1.7		& 0.004		& 0.01 \\
MM6			& 43.2		& 1.9		& 0.004		& 0.01 \\

\hline \

\end{tabular}
\label{CO_param}

\end{center}
\end{table*}

In the {\it ALMA} Band~3 $\rm (84-116\, GHz)$ and Band~7 $\rm (275-373\,GHz)$, we detect three CO transitions, $\rm ^{13}CO\,(1-0)$, $\rm C^{18}O\,(1-0)$ and $\rm C^{17}O\,(3-2)$, with rest frequencies 110.2013543, 109.7821734, and 337.7061123\,GHz, respectively. The rotational transition lines of the isotopologues of CO are well known tracers of both outflow and inflow motion \citep{{1975ApJ...199...79K},{2001ApJ...552L.167Z},{2002A&A...383..892B},{2009ApJ...697L.116W}}. Transitions from different energy levels probe the kinematic structure of different parts of the molecular clouds. Since transitions from higher rotational levels of CO molecules have a higher critical density, their emission arises from high density regions like the dense cores. On the other hand, emission lines corresponding to lower {\it J} transitions trace the kinematics of the low density regions of the molecular cloud \citep{2013A&A...549A...5R}. 

\par Low-resolution spectra probing the $\rm ^{13}CO\,(1-0)$ and $\rm C^{18}O\,(1-0)$ molecular line emission is plotted in {\fig}\ref{co_low}. The beam covers all the cores. Single Gaussians are used to fit these spectra. The $\rm ^{13}CO\,(1-0)$ line peaks at the LSR velocity of the dust clump. Multiple peaks noticeable in $\rm C^{18}O\,(1-0)$ spectrum are the combined effect of the cores and the peak velocities are consistent with the LSR velocities derived for the cores. 
In case of the $\rm C^{17}O\,(3-2)$ line, spectra from the lower resolution Band~7 data cube are extracted over the beam area covering three regions (1) MM1, MM3, and MM4 (2) MM2, and (3) MM5 and MM6 and shown in {\fig}\ref{c17o}(a). Region (1), covering MM1, MM3, and MM4, shows a clear red wing. However, for region (2), probing MM2, the wing is not very prominent. In case of region (3), an inverse P-Cygni profile is seen. A double Gaussian is used to fit this profile. {\fig}\ref{c17o}(b) shows the high-resolution spectra of this line. As discussed earlier, since the separation between MM3 and MM4 is less than the beam size, the spectrum is extracted over a region covering both cores. Given the signal-to-noise ratio of the high-resolution spectra for MM1, MM3, and MM4 it is difficult to interpret the second red-shifted peaks seen as genuine additional components. These features are mostly broad outflow wings. Core MM6 displays a distinct inverse P-Cygni profile and is fitted by a double Gaussian. 

\par Medium resolution $\rm C^{18}O\,(1-0)$ line spectra extracted over the same three regions used for the $\rm C^{17}O\,(3-2)$ spectra are plotted in {\fig}\ref{c18o}(a) with the high-resolution spectra presented in {\fig}\ref{c18o}(b). Similar features are seen though the signal-to-noise of the high-resolution spectra, especially for MM5 and MM6, is poor compared to that of the $\rm C^{17}O\,(3-2)$ line spectra. Also, the medium resolution combined spectrum for MM5 and MM6 shows an additional red-shifted narrow component which is difficult to understand. This component is, however, not detected in the high-resolution spectrum extracted over the same region, and hence could possibly be an artifact.
Line parameters for the three transitions of CO determined by fitting Gaussian profiles using CLASS90 are listed in {\tab}\ref{CO_param}.

\section{DISCUSSION} \label{discussion}

\subsection{Ionized jet}  \label{thermal_jet}

\begin{figure}
\centering
\includegraphics[scale=0.50]{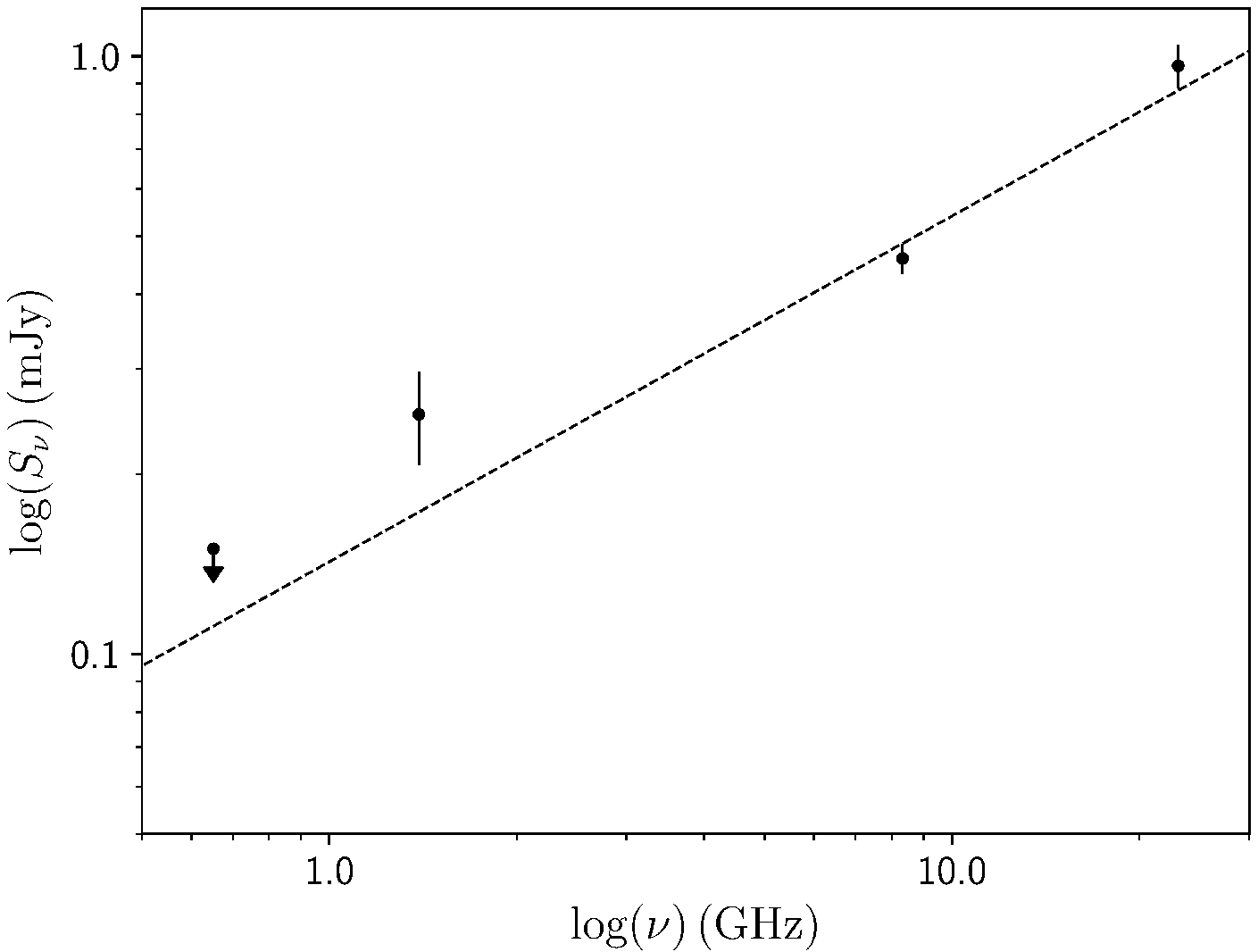}
\caption {Radio SED for component R1 (IRAS\,18264-1152b) using the data points at 1391.6\,MHz, 8\,GHz (3.6\,cm) and 23\,GHz (1.3\,cm). At 651.4~MHz, an upper limit for the flux density is given. The straight line gives the linear fit with spectral index estimate of $0.58\pm0.19$.}
\label{spectral_index}
\end{figure}

Given that {\g19} is classified as an EGO, and hence likely associated with a jet/outflow, we are prompted to investigate the nature of the associated radio emission. As seen in {\fig}\ref{radio_image}(b), the ionized emission associated with {\g19} exhibits a linear morphology consisting of the components R1 and R2 in the east-west direction. The radio emission from both the components is weak ($\rm \sim 0.2\,mJy$) and they are located towards the centroid of a large scale ($\sim$ 1 arcmin) molecular outflow ({\fig}\ref{mom0}a, Section~\ref{outflow}). These traits are in agreement with the typical observational features expected in ionized jets \citep{{1996ASPC...93....3A},{1997IAUS..182...83R},{2016MNRAS.460.1039P},{2019ApJ...880...99R},{2019MNRAS.486.3664O}}.

\par The spectral index is a crucial parameter that helps in understanding the nature of the radiation mechanism. This is defined as $S_\nu \propto \nu^\alpha$, where, $S_\nu$ is the flux density at frequency $\nu$. As discussed in Section \ref{radio_text}, {\g19} is not detected in Band~4. Further, weak emission in Band~5 would render the in-band spectral index estimation highly unreliable as has also been discussed by \citet{2016ApJS..227...25R}.
Emission at 1391.6~MHz encompasses the component IRAS\,18264-1152b. This makes it possible to use this uGMRT sub-band and VLA maps from \citet{2006AJ....131..939Z} to derive the spectral index of R1. An accurate determination of the spectral index requires same spatial scales to be probed at the frequencies used. The synthesized beam sizes of uGMRT 1391.6\,MHz and VLA 8\,GHz (3.6\,cm) maps are nearly identical indicating similar {\it uv} coverage and hence similar spatial scales probed. This is not the case for the 1.3\,cm data. For this, we convolve the map to the match the resolution of the other two frequencies. The radio SED is shown in {\fig}\ref{spectral_index}. For 651.4~MHz, we obtain an upper limit for the flux density ({\it rms} of the map) which is shown in the plot. For the VLA maps, we use the peak flux density as well. The fit to the radio SED of component R1 (IRAS\,18264-1152b) yields a spectral index value of $0.58\pm0.19$. that corroborates well with radio continuum emission originating due to the thermal free-free emission from an ionized collimated stellar wind \citep[and references therein]{2019ApJ...880...99R}. For component R2, similar analysis of spectral index estimation is not possible given that this component is not detected in the VLA maps of \citet{2006AJ....131..939Z}.

\par The spectral index estimation of R1 obtained from the combination of uGMRT and VLA maps of \citet{2006AJ....131..939Z} is in excellent agreement with value of $0.6\pm0.1$ derived by \citet{2016ApJS..227...25R} using flux densities at four central frequencies (4.9 and 7.4\,GHz at 6\,cm and 20.9 and 25.5\,GHz at 1.3\,cm). In comparison, \citet{2006AJ....131..939Z} obtain a lower value of $0.27\pm0.06$. The 3.6\,cm flux density of IRAS\,18264-1152b used by these authors would also include emission from IRAS\,18264-1152a and IRAS\,18264-1152c thus rendering a shallower slope. Based on the weak emission, elongated morphology, association with molecular outflows and rising spectral index values, the radio emission favours shock ionization over the HII region picture. Hence, \citet{2019ApJ...880...99R} list source F (R1) as a ionized jet/wind candidate. Inference of a likely thermal jet or a partially optically thick HII region is drawn by \citet{2006AJ....131..939Z} as well. The identification of R1 as a thermal jet is well supported by the detection of several shock-excited {\h2} lines in the UKIRT-NIR spectra of {\g19} presented in Section~\ref{nir_spec}. To the best of our knowledge, this is the first detection of the ionized jet associated with {\g19} at the lower frequency regime of uGMRT. Similar result is obtained by our group for EGO G12.42+0.50 \citep{2019MNRAS.485.1775I}. The authors propose the coexistence of a thermal jet and a UC {\hii} region driven by the same MYSO to explain the observed radio, mm, and IR emission for G12.42+0.50.

\subsection{A protocluster revealed}

\begin{figure*}
\centering 
\includegraphics[scale=0.30]{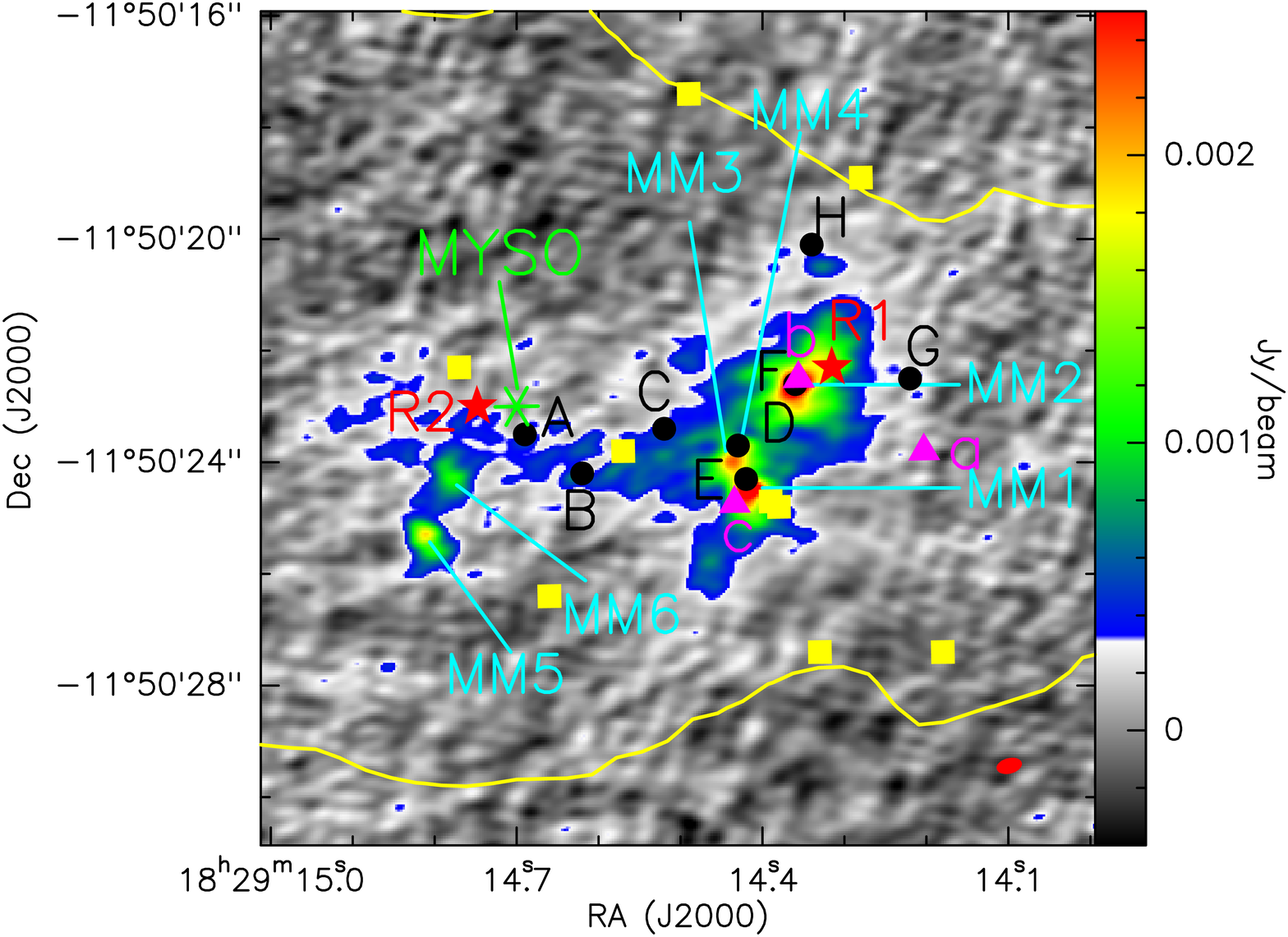}
\caption{Multifrequency picture of the various components of the protocluster. The {\it ALMA} 2.7\,mm map is depicted in the colour scale. The yellow contour traces the $3\sigma$ level of the 4.5{\um} emission. The red stars mark the positions of the radio components R1 and R2 from the uGMRT map. The black filled circles indicate the positions of the radio sources identified by \citet{2016ApJS..227...25R}. The magenta triangles represent the radio components from \citet{2006AJ....131..939Z}. The position of the MYSO associated with {\g19} is indicated with `$\ast$'. The filled yellow squares are the 44\,GHz methanol masers spots in the vicinity of {\g19} identified by \citet{2017ApJS..233....4R}.}
\label{protocluster_image}
\end{figure*}

\begin{table*}
\caption{Identified radio and mm components of the protocluster associated with {\g19}.}
\begin{center}

\begin{tabular}{c c c c c} \hline \hline 

Compact radio sources  	& Radio components 	 & Radio components	& Millimetre cores  \\	
VLA$^1$ & VLA$^2$ & uGMRT 1391.6\,MHz$^3$ & {\it ALMA} 2.7\,mm $^3$ & Remark \\		
\hline
A		& -  & R2 & - & MYSO$^4$/Ionized jet\\
B       & -  & - & - & UC{\hii}/Ionized jet \\
C		& -  & - & - & UC{\hii}/Ionized jet \\
D		& -  & - & MM3 \& MM4 & UC{\hii}/Ionized jet\\
E		& IRAS\,18264-1152c	& - & MM1 & UC{\hii}/Ionized jet\\
F		& IRAS\,18264-1152b	& R1 & MM2 & Ionized jet\\
G		& -  & - & - & UC{\hii}/Ionized jet\\
H		& -  & - & - & UC{\hii}/Ionized jet\\
-		& IRAS\,18264-1152a & - & - &  UC{\hii}/Ionized jet\\
-  		& - & - & MM5 & Dense core\\
-  		& - & - & MM6 & Dense core\\
\hline
\end{tabular}
\label{protocluster_members}
\end{center}
$^1$\citet{2016ApJS..227...25R}; $^2$ \citet{2006AJ....131..939Z}; $^3$This work; $^4$\citet{2010AJ....140..196D} 
\end{table*}


EGO {\g19} has been proposed as an outflow powered by a MYSO \citep{2010AJ....140..196D} and a multiple outflow source \citep{2010MNRAS.404..661V}. Several studies \citep[e.g.][]{{2007AJ....134..346C},{2011ApJ...729..124C}} have shown that MYSOs are actually protoclusters harbouring protostars with different masses and in various stages of evolution. Given the predilection of high-mass stars to form in clustered environments, it is likely that EGOs, in fact harbour protoclusters rather than single high-mass stars. Based on SOFIA FORCAST imaging and archival infrared data of twelve EGOs, \citet{2019ApJ...875..135T} suggest the number of massive sources per EGO to be between 0.9 to 1.9. Similar inferences have been drawn from studies towards other identified EGOs \citep[e.g.][]{{2011ApJ...739L..16B},{2012ApJ...760L..20C}}. 

\par The 870\,{\um} clump cocoons the EGO {\g19} where the extended 4.5\,{\um} emission lies towards the centre as is seen in Figure \ref{870_alma_cores}(a). In {\fig}\ref{protocluster_image}, we present the various radio and mm sources associated with this EGO. The 2.7\,mm dust emission, with the cluster of detected cores lie deeply embedded within the 4.5\,{\um} emission. Portion of the $3\sigma$ contour of the 4.5\,{\um} emission is seen in the region displayed in the figure. Within the dust emission lie the two radio components R1 and R2 mapped at 1391.6~MHz with uGMRT.
As discussed in the previous section, R1 (VLA source F) is most likely an ionized thermal jet. Regarding component R2 (VLA source A) and the other six compact VLA sources (B, C, D, E, G, and H) located within the EGO, \citet{2016ApJS..227...25R} discuss the likelihood of them being {\hii} regions around rapidly accreting massive stars that quench the UV photons thus giving rise to weak radio sources. They also keep alive the debate of these being ionized jets instead. \citet{2006AJ....131..939Z} also suggest that the component IRAS\,18264-1152c (VLA source E) is likely to be an optically thick {\hii} region. These authors detect an additional component IRAS\,18264-1152a of similar nature taking the number of distinct radio sources to be nine. Furthermore, from the analysis of {\it ALMA} 2.7\,mm high-resolution continuum map, we discover six compact cores located within the EGO. Table \ref {protocluster_members} lists the radio and mm sources and their cross-identification.

\par Given the detection of several radio components and mm cores, the picture of a protocluster with multiple (eleven) members is clearly perceivable. The observational signatures of the members show clear evidence of different evolutionary phases. However, high-sensitivity radio surveys of ionized jets have shown association with extended lobes and string of radio knots \citep[e.g.][]{{2003ApJ...587..739G},{2010ApJ...725..734G},{2016MNRAS.460.1039P},{2017ApJ...843...99H}}. From the alignment of the radio sources B, C and F, and the spectral index of F, one can visualize a similar scenario. It is thus plausible that we are probing a string of knots of a single ionized jet driven by R1/MM2/F. Similarly, the components A and G could be part of this jet system though the possibility of another jet driven by R2/MYSO/A cannot be ruled out. If these radio components are indeed knots/lobes of the identified jet-systems, then the number of distinct members of the detected protocluster would be less than eleven. Further studies are required to interpret the exact nature the compact radio sources.

\subsubsection{$CH_3OH$ rotational temperature}

\begin{figure}
\centering 
\includegraphics[scale=0.32]{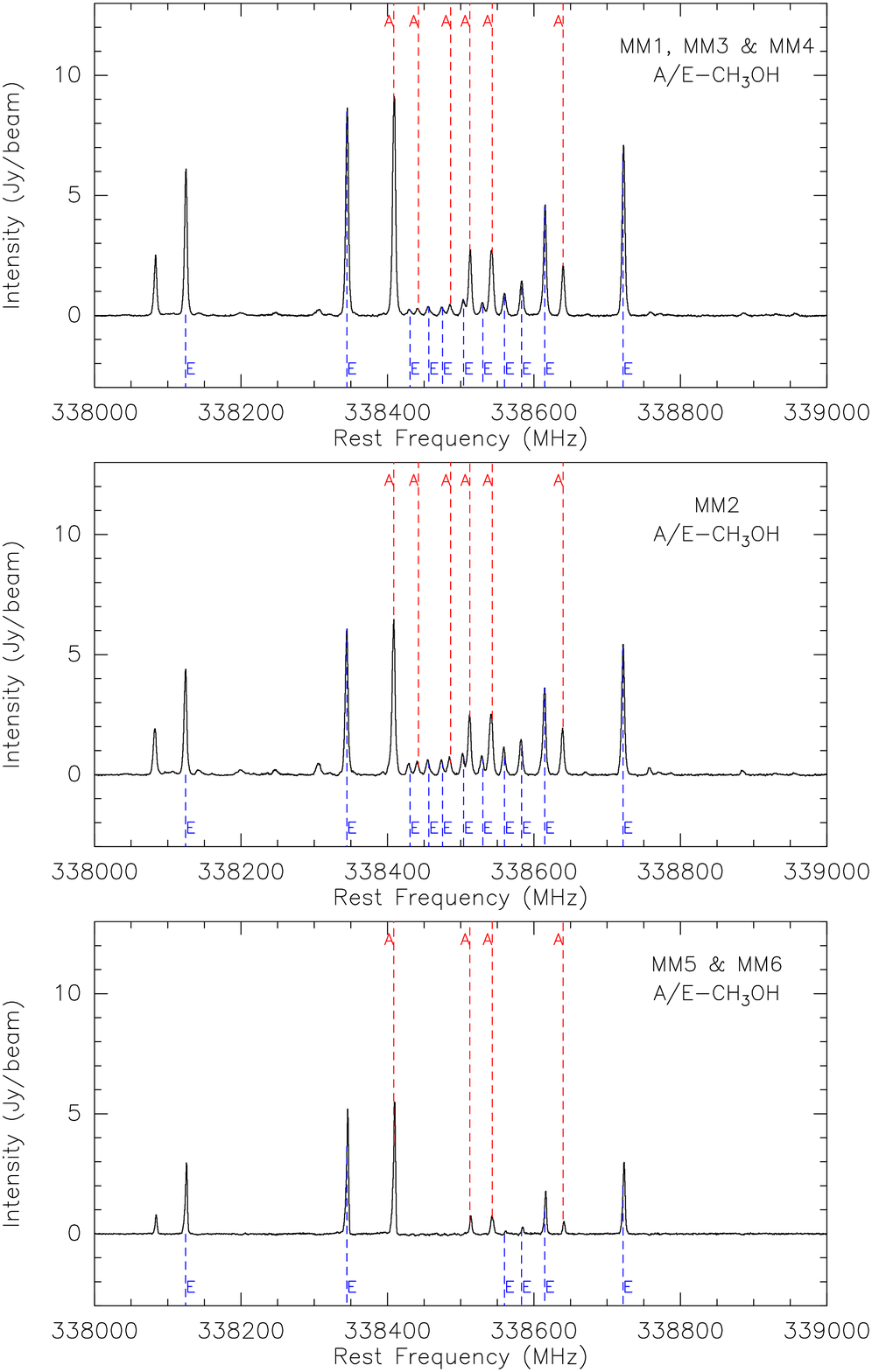}
\caption{Spectra of the $A/E-\rm CH_3OH$ lines detected towards {\g19} extracted over three regions from the low-resolution Band~7 spectral cube. The top panel is the spectrum extracted over the region covering MM1, MM3 and MM4, middle panel corresponds to MM2 and the bottom panel is over the region covering MM5 and MM6. The dashed red and blue lines indicate the peak positions of the $A-\rm CH_3OH$ and $E-\rm CH_3OH$ lines, respectively.}
\label{ch3oh_avg}
\end{figure}

\begin{figure*}
\centering 
\includegraphics[scale=0.19]{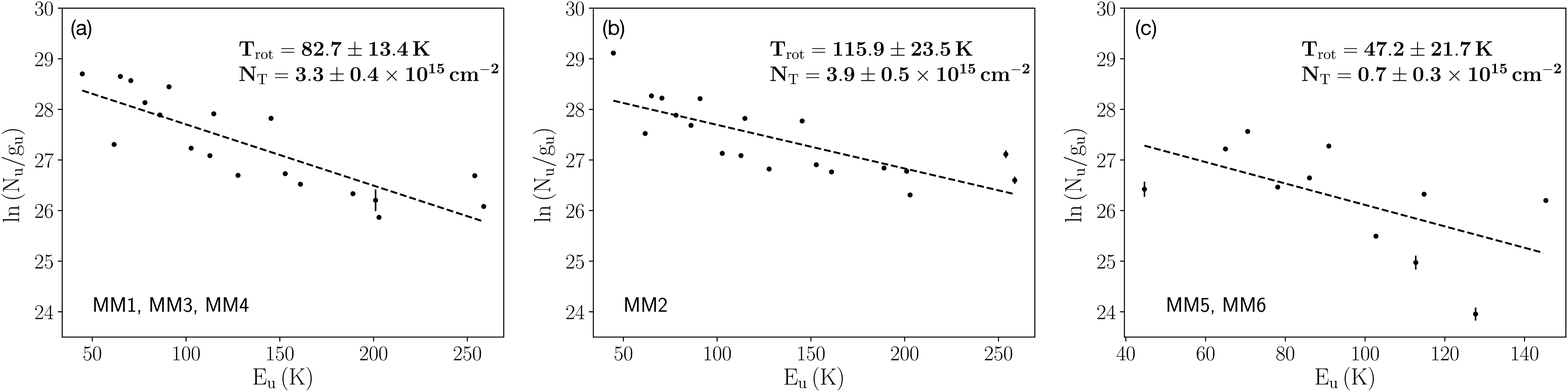} 
\caption{The rotational temperature diagram of the $\rm CH_3OH$ lines at the mm cores, MM1 through MM6 towards {\g19}. The circles indicate the data points with 3$\sigma$ error bars. The linear least-squares fit is indicated by the dashed line.}
\label{rot_diagram}
\end{figure*}

Molecular line transitions can be used to analyse the core properties. 
As seen in {\fig}\ref{ch3oh}, multiple transitions of $\rm CH_3OH$ are detected towards the mm cores associated with {\g19}. This motivates us to use rotational temperature diagram (RTD) (also referred to as Boltzman diagram) to estimate the rotational temperature and the beam averaged column density of the identified mm cores. 
To generate the RTDs, we use the low-resolution (5.3\,arcsec$\times$2.6\,arcsec) ALMA data cubes. Given the beam size, we extract the spectrum covering three regions (1) MM1, MM3, and MM4, (2) MM2, and (3) MM5 and MM6.  The retrieved spectra are shown in {\fig}\ref{ch3oh_avg} and the line parameters are listed in {\tab}\ref{ch3oh_tab_mm1},\ref{ch3oh_tab_mm2}, and \ref{ch3oh_tab_mm5}. The detected methanol lines have linewidths of several $\rm km\,s^{-1}$ and none show unusually high line intensities. Thus these are unlikely to be associated with maser excitations.

\par Considering the molecular emission to be optically thin, the column density of the upper level, $N_{\rm u}$ can be computed from the integrated line intensity following the expression, 
\begin{equation}
N_{\rm u}=\frac{8\pi k \nu^2}{h c^3 A_{\rm u}}\int I_\nu(\textrm {Jy\,beam}^{-1}) dV(\textrm {km\,s}^{-1}) /\eta_{\rm bf}
\end{equation}
\noindent
where $A_{\rm u}$ is Einstein coefficient and $\eta_{\rm bf}$ is the beam-filling factor, taken to be 1 since the emission region is larger than the beam size \citep{2012MNRAS.422.1098R}.
Further, under conditions of local thermodynamic equilibrium,
the measured line intensities are proportional to the level populations that are characterized by a single rotational temperature. The rotational temperature, $T_{\rm rot} $ and the beam averaged column density, $N_{\rm T}$ can be determined using the expression \citep{{1987ApJ...315..621B},{2002ApJ...576..255L},{2004ApJ...617..384R}},
\begin{multline}
\textrm{ln}\bigg(\frac{N_{\rm u}}{g_{\rm u}}\bigg) = \textrm{ln}\bigg(\frac{N_{\rm T}}{Q_{\rm rot}}\bigg)-\frac{E_{\rm u}}{T_{\rm rot}} \\ = \textrm{ln} \bigg[2.04 \times 10^{20} \frac{\int I_\nu(\textrm {Jy\,beam}^{-1}) dV(\textrm {km\,s}^{-1})}{\theta_a \theta_b\,(\textrm {arcsec}^2)\nu^{3}(\rm GHz)S\mu^2(\textrm {debye}^2)}\bigg]
\end{multline}
\noindent
where, $N_{\rm u}$, $g_{\rm u}$ and $E_{\rm u}$ are the upper-level column density, rotational degeneracy factor, and energy, respectively, $Q_{\rm rot} (= 1.23T_{\rm rot}^{1.5})$ is the rotational partition function \citep{{2004ApJ...617..384R},{2012MNRAS.422.1098R}}, $\theta_a$ and $\theta_b$ are the beam sizes, $\nu$ is the rest frequency, $S$ is the line strength, $\mu$ is the dipole moment, and the integration of $I_\nu$ over velocity, $V$, is the integrated line intensity. Values for $E_{\rm u}$ and $S\mu^{2}$ for detected transitions are retrieved from {\it splatalogue - database for astronomical spectroscopy}\footnote{\url{http://www.cv.nrao.edu/php/splat/}} and compiled in {\tab}\ref{ch3oh_par}. The generated RTDs towards the mm cores are plotted in {\fig}\ref{rot_diagram}. Linear least-square fits to the data points gives the estimate of the rotational temperature, $T_{\rm rot}$ and the beam averaged column density, $N_{\rm T}$ of the cores and are listed in {\tab}\ref{mm_params}. For cores MM1, MM3, and MM4, the estimates would be an average value of the three clumps as is the case with the estimates for MM5 and MM6. Estimated values of $T_{\rm rot}$ and $N_{\rm T}$ lie in the range $\sim \rm 47-116\,K$ and $\sim \rm 0.7-4\times 10^{15}\,cm^{-2}$, respectively with core MM2 being the hottest with a rotational temperature of 115.9\,K. The rotational temperatures for the cores are greater than the average cold dust temperature of 18.6\,K deduced for the dust clump (Section~\ref{dust_clump}). The difference in these temperatures could be that the $\rm CH_3OH$ emission arises from the hot cores. The gas and dust could also be thermally decoupled and the higher rotational temperatures could be the result of enhanced collisional excitations of the molecular gas by shocks in the outflow region. A large scatter is seen in the combined RTD of MM5 and MM6 possibly indicating more than one temperature components. However, higher resolution and better signal-to-noise ratio data are required to address this. 

\par To examine the assumption of optically thin transitions, we derived the optical depth of the $\rm CH_3OH$ lines adopting the following expression from \citet{2012MNRAS.422.1098R}
\begin{equation}
\tau=\frac{c^3\sqrt{4\ln2}}{8\pi\nu^3\sqrt{\pi}\Delta V}N_{\rm u}A_{\rm u}\bigg[\exp\bigg(\frac{h\nu}{k T_{\rm rot}}\bigg)-1\bigg]
\end{equation}
The estimated values are listed in {\tab}\ref{ch3oh_tab_mm1},\ref{ch3oh_tab_mm2}, and \ref{ch3oh_tab_mm5}. Only two transitions for the combined cores of MM1, MM3, and MM4 show values $\sim 0.1$. Rest of the lines are far more optically thin.
Considering $T_d$ (18.6 K) instead of $T_{\rm rot}$, to account for temperature uncertainties, if any, also gives optically thin estimates for the transitions. In case of optically thick lines, effects of optical depth can be taken into account by multiplying the term ${N_{\rm u}}/{g_{\rm u}}$ with a correction factor $C_{\tau}=\tau/(1-{\rm e}^{-\tau})$ and subsequently fitting for the rotational temperature iteratively \citep{2012MNRAS.422.1098R}. We also estimate the relative abundance of $\rm CH_3OH$ with respect to the {\h2} molecule and tabulate the values in {\tab}\ref{mm_params} along with the {\h2} number density. 

\par \citet{1986A&A...157..318M}, in their study of $E$-type methanol lines towards a few Galactic molecular line sources concluded that these lines originate from hot clumps embedded in the molecular gas and have densities $\sim \rm 10^6-10^7\,cm^{-3}$, temperatures $\sim \rm 100\,K$, and sizes $\sim \rm 1\,pc$. The abundance relative to the {\h2} molecule was found to be in the range of $10^{-7}-10^{-6}$. In another study of molecular clouds, \citet{2007A&A...466..215L} study the early phases of high-mass stars, particularly the protostellar objects and IRDCs, using methanol as a diagnostic tool.
In their study, they discuss on the origin of $\rm CH_3OH$ lines which could arise from the different components of the protostellar object, which includes the overall emission from the clump and the dense, hot core. In some cases contribution of outflow emission is also included. Temperatures, densities, and abundances fall in the range of $\rm 17-36\,K$, $\rm 2\times10^5-3\times10^6\,cm^{-3}$, and $7\times10^{-10}-2\times10^{-8}$, respectively, if the molecular transition occurs from the bulk emission of the clump. The range changes to $\rm 60-300\,K$, $\rm \gtrsim 10^6 cm^{-3}$, and $\gtrsim 10^{-7}$, respectively for hot cores. The estimated values for cores associated with {\g19} indicates the origin of the $\rm CH_3OH$ emission from hot cores, especially for MM1 to MM4.

\subsubsection{Nature of mm cores} \label{core_props}

\begin{table*}
\caption{Derived parameters of the mm cores associated with {\g19}.}
\begin{center}

\begin{tabular}{l  c c c c c c c} \hline \hline 

Core 	& Radius ($10^{-3}$\,pc)	 & $T_{\rm rot}$ (K)	&$N_{\rm T}\,(\rm 10^{15}\,cm^{-2})$	& $N_{\rm T}/N(\rm H_2)\,(\rm 10^{-7})$	&Mass$_{T_{\rm rot}}$ ($M_\odot$)		  &$n(\rm H_2)\,(\rm 10^{8}\,cm^{-3})$  \\			
\hline \
MM1			&5.4		& 82.7$\pm$13.4		&3.3$\pm$0.4	&0.6	&14.5		&3.2	\\	
MM2 		&10.1		& 115.9$\pm$23.5	&3.9$\pm$0.5	&0.7	&29.6 		&1.0	\\
MM3 \& MM4	&6.6	 	& 82.7$\pm$13.4 	&3.3$\pm$0.4	&0.6	&16.1		&1.9	\\
MM5			&5.2		& 47.2$\pm$21.7		&0.7$\pm$0.3	&0.1	&13.1		&3.2	\\
MM6			&5.3		& 47.2$\pm$21.7		&0.7$\pm$0.3	&0.1	&9.5		&2.2	\\

\hline \
\end{tabular}
\label{mm_params}

\end{center}
\end{table*}

To decipher the nature of the mm cores, we estimate the masses. This is done following the formalism described in \citet{2018ApJ...853..160C}. Assuming optically thin emission, the core masses can be expressed as 
\begin{eqnarray}
  \textrm{M} & = &
  \displaystyle 0.0417 \, M_{\odot}
  \left( {\textrm e}^{0.533 (\lambda / {1.3\, \textrm {mm}})^{-1}
      (T / {20\, \textrm {K}})^{-1}} - 1 \right) \left( \frac{F_{\nu}}{\textrm {mJy}} \right) \nonumber \\
  & & \displaystyle
 \times \left( \frac{\kappa_{\nu}}{0.00638\,\textrm{cm}^2\,\textrm{g}^{-1}} \right)^{-1}
  \left( \frac{d}{2.4\,\textrm {kpc}} \right)^2
  \left( \frac{\lambda}{1.3\, \textrm {mm}} \right)^{3} 
  \label{core_mass}
\end{eqnarray}

\noindent
where the opacity
\begin{equation}
\kappa_{\nu}=0.1({\nu}/\textrm{1000~GHz})^{\beta}\,{\rm cm}^{2}\,{\rm g}^{-1} 
\end{equation}
\noindent
with $\beta$, the dust emissivity spectral index, fixed at 2.0 is the same as used for modelling the FIR SED. $F_\nu$ is the integrated flux density of each component, $d$ is the distance to the source and $\lambda$ is the wavelength taken as 2.7\,mm. Since we do not have the information of dust temperatures at the dense core scales, we use the estimated $T_{\rm rot}$ values in the above expression to derive masses. Compared to the average dust temperature of the clump, these temperatures would be a better choice to represent the hot cores.
The mass estimates range between $10 - 30\,{\rm M_\odot}$ and are listed in {\tab}\ref{mm_params}. Angular sizes given in {\tab}\ref{mm_table} are converted to linear sizes and listed in the table. From the estimated masses and sizes, the detected cores satisfy the criterion, $ m(r) > 870{\rm M_\odot}(r/\rm pc)^{1.33} $ \citep{2010ApJ...716..433K} and qualify as high-mass star forming cores. In comparison to the numbers discussed by \citet{2019ApJ...875..135T}, we detect as many as eleven possible distinct massive components in the {\g19} protocluster that include MYSOs driving ionized jets/outflows and UC{\hii} regions.  

\subsection{Kinematics of the protocluster {\g19}} \label{kinematics}

\subsubsection{Infall signature in MM6} \label{infall}

\begin{table}
\caption{The infall velocity, $V_{\rm inf}$ and mass infall rate, $\dot{M}_{\rm inf}$ of core MM6 associated with {\g19}, estimated using the $\rm C^{17}O\,(3-2)$ line.}
\begin{center}

\begin{tabular}{c c c c c} \hline \hline 

Core  &	$V_{\rm LSR}$ 	&$V_{\rm R}$	&$V_{\rm inf}$   	& $\dot{M}_{\rm inf}$ 	\\

	&($\rm km~s^{-1}$)	&($\rm km~s^{-1}$)	&($\rm km~s^{-1}$)		&($10^{-3}\,M_\odot\,\rm yr^{-1}$)			\\

\hline \
MM6		&42.3	&45.2		& 2.9		& 16.1 \\

\hline \
\end{tabular}
\label{infall_tab}

\end{center}
\end{table}

As discussed in Section\ref{molecular}, CO transitions from higher energy levels are likely tracers of the inner parts of molecular clouds. The $\rm C^{17}O\,(3-2)$ molecular transition shows an inverse P-Cygni profile for core MM6.  
Inverse P-Cygni profiles are generally considered as evidence of gas infall \citep{{2009ApJ...697L.116W},{2013ApJ...776...29L}}. In such profiles, the redshifted absorption feature originates from the dense gas along the line-of-sight and located towards the near side of the observer and moving away.
The emission component is due to the gas on the far side of the central source with its motion towards the observer. Although several studies have attributed the presence of an inverse P-Cygni or redshifted absorption profile to possible gas inflow to the central core \citep[e.g.][]{{1987ApJ...323L.117K},{2008A&A...479L..25Z},{2008ApJ...677..353Q},{2013MNRAS.436.1335L},{2013ApJ...776...29L}}, detailed investigations of inverse P-Cygni profiles of CO lines in high-mass star forming regions are fairly limited \citep[e.g.][]{{2009ApJ...697L.116W},{2011A&A...525A.151B},{2011A&A...530A..53K}}.

\par We make use of the $\rm C^{17}O\,(3-2)$ line profile to estimate the infall velocity and the mass infall rate of the compact core, MM6 associated with {\g19}. The infall velocity is estimated as $V_{\rm inf}$ = $|V_{\rm LSR} - V_{\rm R}|$ \citep{2011A&A...525A.151B}, where $V_{\rm R}$ is the peak of the redshited absorption. From the infall velocity, the mass infall rate is determined following the expression $\dot{M}_{\rm inf}=4\pi R^2V_{\rm inf}\rho$ \citep{2010A&A...517A..66L}, where $\rho$ is the average volume density of the cores, given by $\rho = M/\frac{4}{3} \pi R^3$. From the estimated core mass and radius, the infall velocity and mass infall rate are calculated to be $\rm 2.9\,km\,s^{-1}$ and $16.1\times10^{-3}\,M_\odot\,\rm yr^{-1}$, respectively and listed in {\tab}\ref{infall_tab}. The infall velocity and mass infall rate are consistent with the typical values of several $\rm km\,s^{-1}$ and $\sim 10^{-3}-10^{-2}\,M_\odot\,\rm yr^{-1}$, respectively, found towards hot molecular cores \citep{2013ApJ...776...29L}.

\subsubsection{Outflow activity} \label{outflow}
\begin{figure}
\centering 
\includegraphics[scale=0.23]{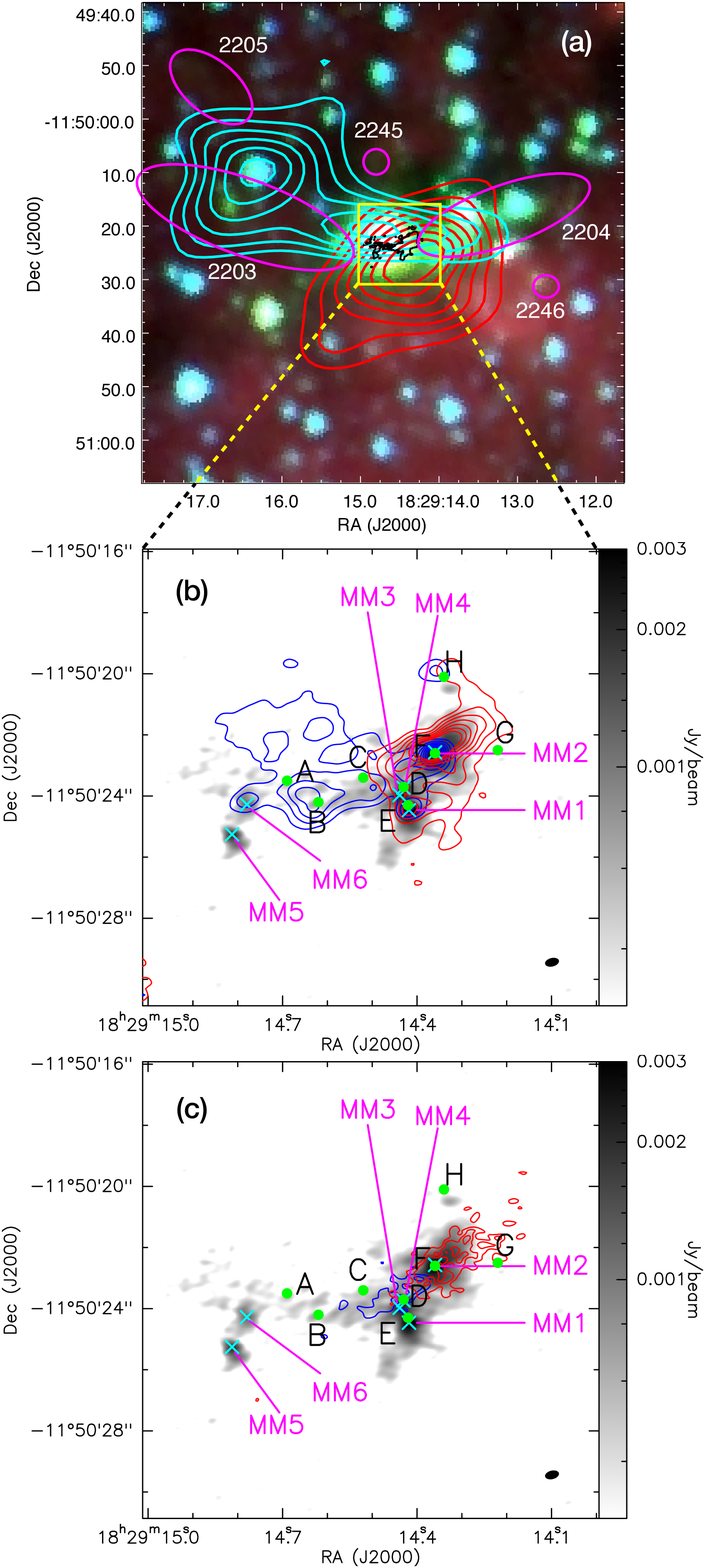}
\caption{(a) The {\it Spitzer} IRAC colour composite image of {\g19} with the 3$\sigma$ ($\rm \sigma = 0.12\,mJy\,beam^{-1}$) {\it ALMA} 2.7\,mm emission contours in black. The moment zero map from the the low-resolution data of the $\rm C^{18}O\,(1-0)$ line integrated from 36.1 to 42.1\,$\rm km\,s^{-1}$ and from 45.1 to 51.1\,$\rm km\,s^{-1}$ are represented using blue and red contours, respectively. The contours start from the 3$\sigma$ level for both the blue and red lobes and increase in steps of 2$\sigma$ ($\sigma=0.1$ (blue); $\rm 0.2\,Jy\,beam^{-1}\,km\,s^{-1}$ (red)). Magenta ellipses denote the location of the {\h2} knots identified by \citet{{2012ApJS..200....2L},{2010MNRAS.404..661V}}. (b) The 2.7\,mm emission is depicted in the grey scale. The positions of the mm cores are marked using `x's. The moment zero map from the high-resolution data of $\rm C^{17}O\,(3-2)$ line over the velocity ranges 36.1 to 42.1\,$\rm km\,s^{-1}$ and 45.1 to 51.1\,$\rm km\,s^{-1}$, covering the blue and red lobes are illustrated in blue and red contours, respectively. The contours star from 5$\sigma$ level for both lobes and increase in steps of 4$\sigma$ ($\rm \sigma = 16.2\,mJy\,beam^{-1}\,km\,s^{-1}$). The filled circles correspond to the radio sources detected by \citet{2016ApJS..227...25R}. (c) Same as (b) for the high-resolution $\rm C^{18}O\,(1-0)$ line integrated over the velocity ranges 41.6 to 43.6\,$\rm km\,s^{-1}$ and 43.6 to 45.6\,$\rm km\,s^{-1}$ covering the blue and red lobes, respectively. The contours star from 3$\sigma$ level for both lobes and increase in steps of 1$\sigma$ ($\rm \sigma = 7.7\,mJy\,beam^{-1}\,km\,s^{-1}$).}
\label{mom0}
\end{figure}

\begin{figure*}
\centering 
\includegraphics[scale=0.8]{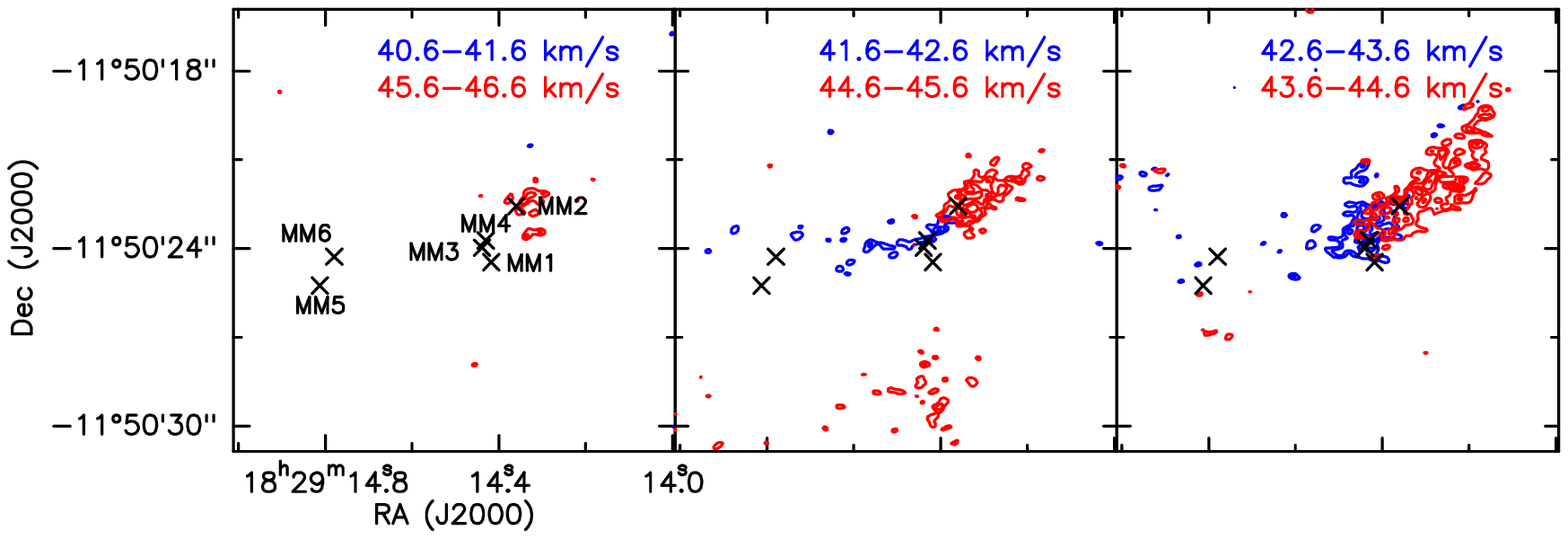}
\caption{Channel maps of $\rm C^{18}O\,(3-2)$ line associated with {\g19} with a velocity resolution of $\rm 1\,km\,s^{-1}$ from the high-resolution data. Each box contains a pair of maps corresponding to the red and blueshifted emissions at the same offset from the LSR velocity. Channel widths are indicated at the top left of each map. The contours start from the 3$\sigma$ level of each map and increases in steps of 1$\sigma$. The positions of MM1, MM2, MM3, MM4, MM5, and MM6 are marked.}
\label{chan_c18o}
\end{figure*}

Co-existing with ionized jets, molecular outflows are ubiquitous in active star forming cores \citep[e.g.][]{{2002A&A...383..892B},{2016MNRAS.460.1039P}}. These outflows are understood to be driven by the jets that entrain the gas and dust from the surrounding clump. Based on a large number of studies, broad wings of CO lines are accepted as signatures of molecular outflow \citep{{2002A&A...383..892B},{2009ApJ...696...66Q}}. The detection of the isotopologues of CO towards {\g19} enables us to investigate the associated outflow(s). 
Towards this, we construct the zeroth moment map of the $\rm C^{18}O\,(1-0)$ emission from the low-resolution cube using the {\small CASA} task, {\tt IMMOMENTS}. The velocity range to integrate is estimated from the extracted spectra discussed earlier and shown in {\fig}\ref{co_low}(b). Following \citet{2014A&A...565A..34S} and \citet{2019MNRAS.485.1775I}, we define the inner limits of the velocity at $\sim V_{\rm LSR}\pm\rm FWHM/2$, where FWHM is the $\rm C^{18}O$ linewidth. Thus, for the blue wing, the emission is integrated over the velocity range of 36.1 to 42.1\,$\rm km\,s^{-1}$. Similarly, for the red wing the integration is done from 45.1 to 51.1\,$\rm km\,s^{-1}$. The spatial extent and morphology of the outflow is revealed in {\fig}\ref{mom0}(a). To facilitate morphological correlation of the outflow with the EGO, the contours of the blue and red lobes are overlaid on the {\it Spitzer} IRAC colour composite image. Magenta ellipses denote the location of the {\h2} knots identified by \citet{2012ApJS..200....2L} and \citet{2010MNRAS.404..661V} from the continuum subtracted {\h2} image. Two distinct and spatially separated outflow lobes in the north-east and south-west direction are identified with the central part of {\g19} located towards the centroid of the outflow. The direction and location of the blue lobe is towards MHOs~2203 and 2205. In their survey of massive molecular outflows in high-mass star forming regions using the $\rm ^{12}CO\,(2-1)$ transition, \citet{2002A&A...383..892B} discuss the detection of a bipolar CO outflow towards the infrared source, IRAS\,18264-1152, associated with {\g19}. The orientation is consistent with that seen in {\fig}\ref{mom0}(a). 

\par The figure also shows the 2.7\,mm emission outlined by the black contour. A zoomed-in picture of the 2.7\,mm map covering the central part of the extended green emission is displayed in {\fig}\ref{mom0}(b) and (c). We investigate outflow signatures in the high-resolution $\rm C^{17}O\,(3-2)$ and $\rm C^{18}O\,(1-0)$ emission in Band~7 and Band~3 cubes, respectively. From a visual inspection of the outflow wings seen in the spectra presented in {\fig}\ref{c17o}, the moment zero maps of the $\rm C^{17}O\,(3-2)$ over the velocity ranges 36.1 to 42.1\,$\rm km\,s^{-1}$ and 45.1 to 51.1\,$\rm km\,s^{-1}$ are constructed. The contours of the moment zero maps are overlaid in {\fig}\ref{mom0}(b). Similarly, for the $\rm C^{18}O\,(1-0)$ emission, the moment zero maps are built over the velocity ranges 41.6 to 43.6\,$\rm km\,s^{-1}$ and 43.6 to 45.6\,$\rm km\,s^{-1}$, the contours of which are overlaid on {\fig}\ref{mom0}(c). Since the high-resolution $\rm C^{18}O\,(1-0)$ data have poor signal-to-noise ratio in comparison to the high-resolution $\rm C^{17}O\,(3-2)$, no emission was detected above the noise beyond the outer limits, 41.6 and 45.6\,$\rm km\,s^{-1}$  of the blue and red lobes, respectively. Interestingly, from the contours of the moment zero maps overlaid in {\fig}\ref{mom0}(b) and (c), the high-resolution data traces another relatively collimated outflow in the south-east and north-west direction with core MM2 located in the close vicinity of its centroid. 

\par To delve further, we generate channel maps of the high-resolution $\rm C^{18}O\,(1-0)$ transition shown in {\fig}\ref{chan_c18o}. The emission in the velocity range $\rm 41.6 - 43.6\,km\,s^{-1}$ is elongated towards south-east of core MM2, whereas the emission with velocities $\rm 44.6 - 45.6\,km\,s^{-1}$, the elongation is towards north-west of MM2 thus giving clear indication of the presence of an outflow. Close to the LSR velocity of the MM2 ($\rm 44.6\,km\,s^{-1}$), as expected,  there is overlap of emission from the blue and the red components. The small projected velocities seen is possibly due to the effect of small inclination angle. 

\par Using the high-resolution $\rm SiO(2-1)$ emission, \citet{2007ApJ...654..361Q} detected the presence of two quasi-perpendicular collimated outflows, one in the south-east and north-west direction and one in the north-east direction. Our high-resolution ALMA map probes the central portion of the outflow presented in \citet{2007ApJ...654..361Q}. A close visual scrutiny of the figures presented by these authors indicate that the
south-east and north-west SiO outflow is aligned with the 
$\rm C^{17}O\,(3-2)$ and $\rm C^{18}O\,(1-0)$ outflows displayed in {\fig}\ref{mom0}(b) and (c). Comparison with the {\h2} outflows and the MHOs show a possible alignment with MHO~2204. 
Further, the direction of north-east $\rm SiO(2-1)$ outflow is consistent with MHO~2245.
SiO is believed to be an excellent tracer of collimated molecular jets and shocks in the interstellar molecular clouds, since it is formed due to the sputtering of Si atoms from the grains due to fast shocks \citep{{2001A&A...372..281H},{2011A&A...526L...2L},{2006A&A...460..721M}}. In studies of several massive star forming regions, protostellar jets were revealed by SiO emission \citep[e.g.][]{{1999A&A...345..949C},{2013A&A...554A..35L},{2013A&A...550A..81C}}. From their study of the massive star formation region IRAS 19410+2336, \citet{2016A&A...589A..29W} conclude that the SiO emission is possible from the high-velocity C-type or CJ-type shocks associated with protostellar jets. uGMRT results have identified an ionized jet with the core MM2 which is in excellent agreement with the collimated outflow from high-resolution $\rm C^{18}O\,(1-0)$ transition and the co-spatial SiO outflow of \citet{2007ApJ...654..361Q}. Several studies point to the possibility that the massive outflows are powered by mm sources \citep[e.g.][]{{2002A&A...383..892B},{2008A&A...488..579M}}. 
From the spatial extent and the maximum velocity of the flow, the dynamical timescale ($T_{\rm dyn}=L_{\rm flow}/v_{\rm max}$) is estimated to be of the order of $\rm 10^4\,yr$ which is typical of outflows originating from mm cores rather than UC {\hii} regions \citep{{1998ApJ...507..861S},{2003ApJ...584..882S},{2009ApJ...696...66Q}}. This lends crucial support to our conclusion of R1 (MM2) being an ionized jet.

\par Based on multiwavelength data, the results and discussion presented in this paper enables us to draw a comprehensive picture of the star forming region associated with {\g19}. The EGO, which is embedded in a massive cold clump within an IRDC, is unveiled to be a young and active protocluster harbouring several massive components in early evolutionary phases. In conjunction with ionized and {\h2} jets, the kinematic picture reveals multiple molecular outflows driven by one or more members of the protocluster.

\section{SUMMARY} \label{summary}
From the multi-frequency and multi-phase study of the EGO, {\g19} an interesting picture of it being a protocluster is revealed with multiple components spanning a wide evolutionary spectrum from hot cores in accretion phase to cores driving multiple outflows to possible UC {\hii} regions. The main results are summarized below:

\begin{enumerate}
\item {\g19} is located  within a massive dust clump of mass 1911\,$M_\odot$, identified from the 870\,{\um} map with the 4.5\,{\um} emission towards its centre. Deeply embedded within this are nine distinct radio components and six dense mm cores unraveling a protocluster. 

\item Deep uGMRT observations detect weak radio emission associated with {\g19}. The radio continuum emission at 1391.6\,MHz exhibits a linear structure consisting of two radio components, R1 and R2, in the east-west direction. The spectral index estimate of the radio component, R1, $0.58\pm0.19$, confirms it as a thermal jet.

\item The shock-excited {\h2} emission is established from the NIR spectroscopy in concordance with the jet/outflow scenario.

\item Six dense and compact dust cores (MM1-MM6) are identified from the high resolution {\it ALMA} map at 2.7\,mm map. The brightest of these cores, MM2, is coincident with the radio component R1. Based on the mass and size estimates, the cores qualify as high-mass star forming cores.

\item RTDs from detected $\rm CH_3OH$ transitions yield rotational temperature and beam averaged column densities that are consistent with hot cores.

\item The $J=3-2$ transition of $\rm C^{17}O$ towards MM6 show a clear inverse P-Cygni profile which is a signature of protostellar infall. 

\item From the $J=1-0$ transition of the $\rm C^{18}O$, we detect the presence of a large scale molecular outflow in the direction of one of the lobes of the bipolar {\h2} outflow. At higher resolution an additional collimated, bipolar outflow in the south-east and north-west direction is detected likely powered by MM2. This suggests that {\g19} harbours multiple outflows.

\end{enumerate}

\section*{ACKNOWLEDGEMENTS}
We thank the referee for valuable suggestions which has helped improve the manuscript. The authors would like to thank Prof. Luis A. Zapata for kindly providing the fits images of the VLA maps of {\g19} presented in \citet{2006AJ....131..939Z}. CHIC acknowledges the support of the Department of Atomic Energy, Government of India, under the project  12-R\&D-TFR-5.02-0700. We thank the staff of the GMRT, who made these observations possible. GMRT is run by the National Centre for Radio Astrophysics of the Tata Institute of Fundamental Research. We also thank the staff of UKIRT for their assistance in the observations. UKIRT is owned by the University of Hawaii (UH) and operated by the UH Institute for Astronomy. When some of the data reported here were acquired, UKIRT was supported by NASA and operated under an agreement among the University of Hawaii, the University of Arizona, and Lockheed Martin Advanced Technology Center; operations were enabled through the cooperation of the East Asian Observatory. This paper makes use of the following ALMA data: ADS/JAO.ALMA$\#$2017.1.00377.S. ALMA is a partnership of ESO (representing its member states), NSF (USA) and NINS (Japan), together with NRC (Canada), MOST and ASIAA (Taiwan), and KASI (Republic of Korea), in cooperation with the Republic of Chile. The Joint ALMA Observatory is operated by ESO, AUI/NRAO and NAOJ. This research made use of NASA/IPAC Infrared Science Archive, which is operated by the Jet Propulsion Laboratory, Caltech under contract with NASA. This publication also made use of data products from {\it Hersche}l (ESA space observatory) and the ATLASGAL data products. The ATLASGAL project is a collaboration between the Max-Planck-Gesellschaft, the European Southern Observatory (ESO) and the Universidad de Chile. 

\bibliography{reference}
\section*{APPENDIX}
\appendix
\renewcommand{\thetable}{A\arabic{table}}
\begin{table*}
\caption{Spectroscopic parameters of the $A-\rm CH_3OH$ and $E-\rm CH_3OH$ lines detected towards {\g19} taken from the spectroscopic databases, CDMS and JPL from {\it spalatalogue}.}
\begin{center}
\begin{tabular}{l l c c} \hline \hline \
Transition		& Frequency 	&$E_u$  	&$S\mu^2$  \\
$J_k$			& (MHz)	 &(K)	&(debye$^2$)		\\
\hline \
$A-\rm CH_3OH$ \\
$7_0-6_0$ 	& 338408.698		&64.98	&5.66		\\
$7_6-6_6$	& 338442.367		&258.7	&3.01	 \\
$7_5-6_5$	& 338486.322		&202.88	&5.57	\\
$7_4-6_4$	& 338512.632		&145.33	&3.82	\\
$7_3-6_3$	& 338543.152		&114.79	&4.61	 \\
$7_2-6_2$	& 338639.802		&102.72	&5.23	 \\
$2_2-3_1$	& 340141.143		&44.67	&0.31 	\\

\hline \
$E-\rm CH_3OH$ \\
$3_3-4_2$ 		& 337135.853			&61.64	&0.99		\\
$7_0-6_0$ 		& 338124.488			&78.08	&5.65		\\
$7_{-1}-6_{-1}$	& 338344.588		&70.55	&5.55		 \\
$7_6-6_6$ 		& 338430.981			&253.95	&1.50		\\
$7_{-5}-6_{-5}$	& 338456.521 	&189.00	&2.76		\\
$7_5-6_5$		& 338475.217			&201.06	&2.76		\\
$7_{-4}-6_{-4}$	& 338504.065		&152.89	&3.81		\\
$7_4-6_4$		& 338530.256			&160.99	&3.82		\\
$7_{-3}-6_{-3}$	& 338559.963		&127.71	&4.64		\\
$7_3-6_3$		& 338583.216			&112.71	&4.63		\\
$7_1-6_1$		& 338614.936			&86.05	&5.68		 \\
$7_2-6_2$		& 338721.693			&87.26	&5.14		\\

\hline \
\end{tabular}
\label{ch3oh_par}
\end{center}
\end{table*}

\begin{table*}
\caption{$A-\rm CH_3OH$ and $E-\rm CH_3OH$ lines towards MM1, MM3 and MM4 of {\g19}. Columns 1 and 2 are the transitions and the corresponding frequencies, respectively. Columns 3-5 are the parameters and the uncertainties from the Gaussian fits to each line; they are line flux ($\int I_\nu dV$), LSR velocity ($V_{\rm LSR}$), and the line width ($\Delta V$). Column 6 is the optical depth for each transition with two values, the first for $T_{\rm rot}$ and the second for $T_d=18.6\,\rm K$, respectively.}
\begin{center}
\begin{tabular}{l l c c c c} \hline \hline \
Transition		& Frequency 	& $V_{\rm LSR}$			& $\int I_\nu dV$ 			& $\Delta V$	& $\tau_{T_{\rm rot}}/\tau_{T_d}$	\\
$J_k$			& (MHz)	& ($\rm km\,s^{-1}$)	& ($\rm Jy\,beam^{-1}\,km\,s^{-1}$)	& ($\rm km\,s^{-1}$) \\
\hline \
$A-\rm CH_3OH$ \\
$7_0-6_0$ 	& 338408.698		&43.6$\pm$0.4		&40.3$\pm$0.9	&4.1$\pm$0.4		& 0.112/0.496	\\
$7_6-6_6$	& 338442.367		&44.9$\pm$0.1		& 1.6$\pm$0.1	&5.6$\pm$0.3		& 0.003/0.015 \\
$7_5-6_5$	& 338486.322		&44.7$\pm$0.1		& 2.5$\pm$0.1	&5.4$\pm$0.2  	& 0.005/0.023 \\
$7_4-6_4$	& 338512.632	 	&43.7$\pm$0.1		&11.9$\pm$0.5	&4.2$\pm$0.2  	& 0.032/0.142 \\
$7_3-6_3$	& 338543.152 	&44.8$\pm$0.4		&15.7$\pm$0.4	&5.4$\pm$0.4  	& 0.033/0.147 \\
$7_2-6_2$	& 338639.802	 	&43.9$\pm$0.4  	&9.0$\pm$0.6		&4.3$\pm$0.4  	& 0.024/0.106 \\
$2_2-3_1$	& 340141.143	  	&44.3$\pm$0.03	&2.4$\pm$0.1		&4.7$\pm$0.1  	& 0.006/0.025 \\

\hline \
$E-\rm CH_3OH$ \\
$3_3-4_2$ 		& 337135.853	 		& 44.6$\pm$0.1	& 1.8$\pm$0.05 	&5.0$\pm$0.1  	& 0.004/0.016 \\
$7_0-6_0$ 		& 338124.488 		& 43.6$\pm$0.01	& 23.9$\pm$0.2	&3.7$\pm$0.03	& 0.073/0.324 \\
$7_{-1}-6_{-1}$	& 338344.588	 	& 43.6$\pm$0.01 & 36.4$\pm$0.2	&3.9$\pm$0.02 	& 0.105/0.466 \\
$7_6-6_6$ 		& 338430.981	 		& 44.9$\pm$0.1	& 1.5$\pm$0.1 		&6.0$\pm$0.3 	& 0.003/0.013 \\
$7_{-5}-6_{-5}$	& 338456.521 	& 44.8$\pm$0.2	& 1.9$\pm$0.2		&5.0$\pm$0.6 	& 0.004/0.019 \\
$7_5-6_5$		& 338475.217	 		& 44.8$\pm$1.1	& 1.7$\pm$0.8		&4.7$\pm$2.1  	& 0.004/0.018 \\
$7_{-4}-6_{-4}$	& 338504.065		& 44.0$\pm$0.1	& 4.0$\pm$0.2		&6.1$\pm$0.4  	& 0.007/0.033 \\
$7_4-6_4$		& 338530.256			& 44.3$\pm$0.2	& 3.2$\pm$0.3		&6.2$\pm$0.7  	& 0.006/0.026 \\
$7_{-3}-6_{-3}$	& 338559.963		& 44.1$\pm$0.1	& 4.7$\pm$0.3		&4.9$\pm$0.3  	& 0.011/0.048 \\
$7_3-6_3$		& 338583.216			& 43.9$\pm$0.1	& 6.9$\pm$0.3		&4.7$\pm$0.2  	& 0.017/0.075 \\
$7_1-6_1$		& 338614.936			& 43.7$\pm$0.4	& 18.9$\pm$0.5		&3.9$\pm$0.4  	& 0.055/0.246 \\
$7_2-6_2$		& 338721.693			& 44.4$\pm$0.4	& 30.3$\pm$0.7 	&4.0$\pm$0.4  	& 0.085/0.380 \\

\hline \
\end{tabular}
\label{ch3oh_tab_mm1}
\end{center}
\end{table*}

\begin{table*}
\caption{Same as {\tab}\ref{ch3oh_tab_mm1} for MM2 towards {\g19}.}
\begin{center}
\begin{tabular}{l l c c c c} \hline \hline \
Transition		& Frequency 	& $V_{\rm LSR}$			& $\int I_\nu dV$ 			& $\Delta V$ & $\tau_{T_{\rm rot}}/\tau_{T_d}$ \\
 (MHz)	& ($\rm km\,s^{-1}$)	& ($\rm Jy\,beam^{-1}\,km\,s^{-1}$)	& ($\rm km\,s^{-1}$) \\
\hline \
$A-\rm CH_3OH$ \\
$7_0-6_0$ 	& 338408.698		&44.1$\pm$0.02	&27.5$\pm$0.2	&4.1$\pm$0.04	& 0.054/0.335 \\
$7_6-6_6$	& 338442.367		&45.6$\pm$0.4		& 2.8$\pm$0.5	&5.0$\pm$1.1  	& 0.004/0.028 \\
$7_5-6_5$	&	338486.322	&45.4$\pm$0.2		& 3.8$\pm$0.3	&5.2$\pm$0.4 	& 0.006/0.037 \\
$7_4-6_4$	& 338512.632		&44.4$\pm$0.1		&11.3$\pm$0.6	&4.3$\pm$0.3  	& 0.021/0.131 \\
$7_3-6_3$	& 338543.152		&45.5$\pm$0.2		&14.3$\pm$0.9	&5.4$\pm$0.4  	& 0.022/0.135 \\
$7_2-6_2$	& 338639.802		&44.5$\pm$0.02  	& 8.2$\pm$0.1	&4.0$\pm$0.1		& 0.016/0.102 \\
$2_2-3_1$	& 340141.143	 	&45.3$\pm$0.4		& 3.6$\pm$0.1	&5.0$\pm$0.4		& 0.006/0.036 \\

\hline \
$E-\rm CH_3OH$ \\
$3_3-4_2$ 		& 337135.853			& 45.5$\pm$0.1		& 2.3$\pm$0.1 	&4.4$\pm$0.1		&   0.004/0.026	\\
$7_0-6_0$ 		& 338124.488			& 44.3$\pm$0.4		&18.7$\pm$0.4	&4.2$\pm$0.4		&	0.036/0.227 \\
$7_{-1}-6_{-1}$	& 338344.588		& 44.2$\pm$0.01	&25.7$\pm$0.1	&4.1$\pm$0.02	&	0.051/0.316 \\
$7_6-6_6$ 		& 338430.981			& 45.7$\pm$0.4		& 2.3$\pm$0.4 	&4.7$\pm$0.9		&	0.004/0.025 \\
$7_{-5}-6_{-5}$	& 338456.521 	& 45.3$\pm$0.4		& 3.2$\pm$0.1	&5.0$\pm$0.4		&	0.005/0.032 \\
$7_5-6_5$		& 338475.217			& 45.3$\pm$0.2		& 3.0$\pm$0.2	&4.7$\pm$0.4		&	0.005/0.032  \\
$7_{-4}-6_{-4}$	& 338504.065		& 45.0$\pm$0.2		& 4.7$\pm$0.4	&5.3$\pm$0.5		&	0.007/0.045 \\
$7_4-6_4$		& 338530.256			& 45.2$\pm$0.1		& 4.1$\pm$0.1	&5.2$\pm$0.2		&	0.006/0.040 \\
$7_{-3}-6_{-3}$	& 338559.963		& 44.9$\pm$0.2		& 5.3$\pm$0.5	&4.5$\pm$0.5		&	0.009/0.059 \\
$7_3-6_3$		& 338583.216			& 44.8$\pm$0.1		& 6.9$\pm$0.2	&4.6$\pm$0.2		&	0.012/0.076  \\
$7_1-6_1$		& 338614.936			& 44.4$\pm$0.4		&15.4$\pm$0.4	&4.0$\pm$0.4		&	0.031/0.193  \\
$7_2-6_2$		& 338721.693			& 44.9$\pm$0.01	&23.9$\pm$0.1 	&4.3$\pm$0.01	&	0.045/0.280 \\

\hline \
\end{tabular}
\label{ch3oh_tab_mm2}
\end{center}
\end{table*}

\begin{table*}
\caption{Same as {\tab}\ref{ch3oh_tab_mm1} for MM5 and MM6 towards {\g19}.}
\begin{center}
\begin{tabular}{l l c c c c} \hline \hline \
Transition		& Frequency 	& $V_{\rm LSR}$			& $\int I_\nu dV$ 			& $\Delta V$ & $\tau_{T_{\rm rot}}/\tau_{T_d}$ \\
 (MHz)	& ($\rm km\,s^{-1}$)	& ($\rm Jy\,beam^{-1}\,km\,s^{-1}$)	& ($\rm km\,s^{-1}$) \\
\hline \
$A-\rm CH_3OH$ \\
$7_0-6_0$ 	& 338408.698		&42.7$\pm$0.01	&9.6$\pm$0.6		&2.1$\pm$0.04 	&   0.091/0.230	\\
$7_4-6_4$	& 338512.632		&42.8$\pm$0.04	&2.3$\pm$0.1		&3.1$\pm$0.1  	&   0.015/0.038	\\
$7_3-6_3$	& 338543.152		&44.0$\pm$0.05	&3.2$\pm$0.1		&4.2$\pm$0.1		&   0.015/0.039	\\
$7_2-6_2$	& 338639.802		&42.9$\pm$0.02	&1.6$\pm$0.03	&2.9$\pm$0.1		&	0.011/0.027 \\
$2_2-3_1$	& 340141.143		&43.0$\pm$0.1		&0.2$\pm$0.02	&2.5$\pm$0.3		&	0.002/0.005 \\

\hline \
$E-\rm CH_3OH$ \\
$7_0-6_0$ 		& 338124.488			& 42.8$\pm$0.01	& 4.5$\pm$0.2	&2.1$\pm$0.04  &	0.043/0.109 \\
$7_{-1}-6_{-1}$	& 338344.588		& 42.9$\pm$0.01 	&13.3$\pm$0.1	&2.5$\pm$0.02 	&	0.107/0.271 \\
$7_{-3}-6_{-3}$	& 338559.963		& 42.5$\pm$0.1		& 0.3$\pm$0.04	&2.3$\pm$0.4		&	0.003/0.007 \\
$7_3-6_3$		& 338583.216			& 42.7$\pm$0.4		& 0.8$\pm$0.3	&2.7$\pm$1.0		&	0.006/0.015 \\
$7_1-6_1$		& 338614.936			& 43.1$\pm$0.1		& 5.5$\pm$0.2	&3.0$\pm$0.1  	&	0.036/0.092 \\
$7_2-6_2$		& 338721.693			& 43.7$\pm$0.02	& 9.4$\pm$0.1 	&3.0$\pm$0.03 	&	0.062/0.158 \\

\hline \
\end{tabular}
\label{ch3oh_tab_mm5}
\end{center}
\end{table*}



\end{document}